\documentclass[epj]{svjour}
\usepackage[sort&compress,numbers]{natbib}
\usepackage{amssymb}
\usepackage{graphicx}
\usepackage{geometry}
\usepackage[dvipsnames,table]{xcolor}
\usepackage{slashed}
\usepackage{hyperref} 
\usepackage{xspace}
\usepackage[tight]{subfigure}
\usepackage{amsmath}
\usepackage{amsfonts}
\usepackage{mathrsfs}
\usepackage{comment}
\usepackage{afterpage}
\usepackage{verbatim}
\usepackage{booktabs}
\usepackage{array}
\usepackage[title,titletoc]{appendix}
\usepackage{tabularx}

\newcolumntype{C}[1]{>{\centering\arraybackslash}p{#1}}

\def\refeq#1{\mbox{(\ref{#1})}}
\def\reffi#1{\mbox{Figure~\ref{#1}}}
\def\reffis#1{\mbox{Figures~\ref{#1}}}
\def\refta#1{\mbox{Table~\ref{#1}}}

\def\refse#1{\mbox{Section~\ref{#1}}}

\def\citere#1{\mbox{Ref.~\cite{#1}}}
\def\citeres#1{\mbox{Refs.~\cite{#1}}}

\newcommand{\newc}{\newcommand}
\newc{\beq}{\begin{equation}}
\newc{\eeq}{\end{equation}}
\newc{\bit}{\begin{itemize}}
\newc{\eit}{\end{itemize}}
\newc{\ben}{\begin{enumerate}}
\newc{\een}{\end{enumerate}}
\newc{\bce}{\begin{center}}
\newc{\ece}{\end{center}}
\newc{\bfi}{\begin{figure}}
\newc{\efi}{\end{figure}}

\newcommand{\rT}{{\mathrm{T}}}
\newcommand{\rR}{{\mathrm{R}}}

\newcommand{\ie}{\emph{i.e.}\ }
\newcommand{\eg}{\emph{e.g.}\ }

\newcommand{\GeV}{\ensuremath{\,\text{GeV}}\xspace}
\newcommand{\TeV}{\ensuremath{\,\text{TeV}}\xspace}
\newcommand{\fb}{{\ensuremath\unskip\,\text{fb}}\xspace}

\newcommand{\PH}{\ensuremath{\text{H}}\xspace}
\newcommand{\Pj}{\ensuremath{\text{j}}\xspace}

\newcommand{\Pe}{\ensuremath{\text{e}}\xspace}
\newcommand{\Pb}{\ensuremath{\text{b}}\xspace}

\newcommand{\Pt}{\ensuremath{\text{t}}\xspace}

\newcommand{\PW}{\ensuremath{\text{W}}\xspace}
\newcommand{\PZ}{\ensuremath{\text{Z}}\xspace}

\newcommand{\bj}{\Pj_{\Pb}\xspace}
                                    
\newcommand{\Mt}{\ensuremath{m_\Pt}\xspace}
\newcommand{\MH}{\ensuremath{M_\PH}\xspace}
\newcommand{\MWOS}{\ensuremath{M_\PW^\text{OS}}\xspace}
\newcommand{\MW}{\ensuremath{M_\PW}\xspace}
\newcommand{\MZOS}{\ensuremath{M_\PZ^\text{OS}}\xspace}
\newcommand{\MZ}{\ensuremath{M_\PZ}\xspace}

\newcommand{\Gt}{\ensuremath{\Gamma_\Pt}\xspace}
\newcommand{\GH}{\ensuremath{\Gamma_\PH}\xspace}

\newcommand{\GZOS}{\ensuremath{\Gamma_\PZ^\text{OS}}\xspace}

\newcommand{\GWOS}{\ensuremath{\Gamma_\PW^\text{OS}}\xspace}

\newcommand{\GF}{\ensuremath{G_\mu}}
\newcommand{\alphas}{\ensuremath{\alpha_\text{s}}\xspace}

\newcommand{\order}[1]{\ensuremath{\mathcal{O}{\left(#1\right)}}\xspace}

\newcommand{\recola}{{\sc Recola}\xspace}

\newcommand{\mocanlo}{{\sc MoCaNLO}\xspace}
\newcommand{\collier}{{\sc Collier}\xspace}

\newcommand{\madgraphnlo}{{\sc\small MadGraph5\_aMC@NLO}\xspace}

\newcolumntype{.}{D{.}{.}{-1}}
\newcolumntype{d}[1]{D{.}{.}{#1}}
\colorlet{tableoverheadcolor}{gray!37.5}
\colorlet{tableheadcolor}{gray!25}
\colorlet{tablerowcolor}{gray!12.5}

\def\draftdate{\relax}
\def\mda{\relax}
\def\mua{\relax}
\def\mla{\relax}
\def\draft{
\def\thtystars{******************************}
\def\sixtystars{\thtystars\thtystars}
\typeout{}
\typeout{\sixtystars**}
\typeout{* Draft mode!
         For final version remove \protect\draft\space in source file *}
\typeout{\sixtystars**}
\typeout{}
\def\draftdate{\today}
\def\mua{\marginpar[\boldmath\hfil$\uparrow$]%
                   {\boldmath$\uparrow$\hfil}\color{black}%
                    \typeout{marginpar: $\uparrow$}\ignorespaces}
\def\mda{\color{red}\marginpar[\boldmath\hfil$\downarrow$]%
                   {\boldmath$\downarrow$\hfil}%
                    \typeout{marginpar: $\downarrow$}\ignorespaces}
\def\mla{\marginpar[\boldmath\hfil$\rightarrow$]%
                   {\boldmath$\leftarrow $\hfil}%
                    \typeout{marginpar: $\leftrightarrow$}\ignorespaces}
\def\Mua{\marginpar[\boldmath\hfil$\Uparrow$]%
                   {\boldmath$\Uparrow$\hfil}\color{black}%
                    \typeout{marginpar: $\uparrow$}\ignorespaces}
\def\Mda{\color{red}\marginpar[\boldmath\hfil$\Downarrow$]%
                   {\boldmath$\Downarrow$\hfil}%
                    \typeout{marginpar: $\downarrow$}\ignorespaces}
\def\Mla{\marginpar[\boldmath\hfil\textcolor{red}{$\Rightarrow$}]%
                   {\boldmath\textcolor{red}{$\Leftarrow $}\hfil}%
                    \typeout{marginpar: $\leftrightarrow$}\ignorespaces}
\overfullrule 5pt
\oddsidemargin 15mm
\marginparwidth 29mm
}

\newcommand{\mc}{\mathcal}
\newcommand{\as}{\alpha_{\textrm{s}}}
\newcommand{\pt}[1]{p_{\rT,{#1}}}
\newcommand{\nnb}{\nonumber}

\newcommand{\eettll}{\Pe^+\Pe^-\to\bj\,\bj\,\ell^-\ell'^+\nu_{\bar \ell}\nu_{\ell'}}
\newcommand{\eettsl}{\Pe^+\Pe^-\rightarrow \bj\, \bj\, \Pj\, \Pj \, \mu^+ \nu_\mu}

\begin{document}
\title{
  NLO QCD corrections to off-shell top--antitop~production with semi-leptonic decays at lepton colliders \\[-4cm]
  \hspace*{11.8cm}\mbox{\small {FR-PHENO-2023-01\quad MPP-2023-28}}\\[3cm]
}

\author{Ansgar Denner\footnote{\label{a}} \and Mathieu Pellen\footnote{\label{b}} \and Giovanni Pelliccioli\footnote{}}
\institute{
  \textsuperscript{a} University of W\"urzburg, Institut f\"ur Theoretische Physik und Astrophysik, Emil-Hilb-Weg 22, 97074 W\"urzburg, Germany\\
  \textsuperscript{b} Universit\"at Freiburg, Physikalisches Institut, D-79104 Freiburg, Germany\\
  \textsuperscript{c} Max-Planck-Institut f\"ur Physik, F\"ohringer Ring 6, 80805 M\"unchen, Germany
}

\abstract{
The study of top-quark properties will be a central aspect of the physics programme of any future lepton collider.
In this article, we investigate the production of top-quark pairs in
the semi-leptonic decay channel in $\rm e^+e^-$ collisions, 
whose experimental signature is one charged lepton, jets, and missing energy.
We present for the first time fiducial cross sections and differential distributions
at next-to-leading-order accuracy in QCD for the full off-shell process.
We find that the QCD corrections for the considered process are strongly dependent on the beam energies and range from few per cent up to more than $100\%$ (near threshold and above $1\TeV$).
We focus, in particular, on two scenarios:
one close to threshold ($365\GeV$), dominated by top-pair production, and
one at the TeV scale ($1.5\TeV$), for which irreducible-background contributions become relevant.
An assessment of polarised-beam effects is also provided.}
\authorrunning{\emph{
  Denner \and Pellen \and Pelliccioli
}}
\titlerunning{\emph{
  NLO QCD corrections to off-shell $\Pt\bar\Pt$ with semi-leptonic decays at lepton colliders
}}

\maketitle
\tableofcontents
\section{Introduction}\label{sec:intro}

What large-scale collider experiment will come after the end of the Large Hadron Collider~(LHC) is currently an open question.
At present, several options are being considered which include lepton colliders such as the International Linear Collider~(ILC)~\cite{ILC:2013jhg, Behnke:2013lya,Bambade:2019fyw}, the FCC-ee \cite{FCC:2018evy}, or the Compact Linear Collider~(CLIC)~\cite{Linssen:2012hp}.
In all cases, the study of top-quark properties will play a central role in the physics programme of those facilities.

The main advantage of lepton colliders over hadron ones is the
possibility to tune very precisely the centre-of-mass (CM) energy of the experiment.
Thus, one can perform a scan of energies covering the threshold region
for the production of a pair of top quarks.
This provides a very clean access to key properties of the top quark such as its mass and width~\cite{Seidel:2013sqa}.

On the experimental side, significant prospective work has been done~\cite{Boronat:2016tgd,CLICdp:2018esa,Dannheim:2019rcr} to estimate the potential gain in performing such measurements and to assess their experimental limitations.
On the theory side, great efforts have been put in providing precise predictions using non-relativistic QCD and resummation techniques at threshold~\cite{Hoang:2010gu,Hoang:2013uda,Beneke:2015kwa,Beneke:2017rdn}.
Differential predictions including also the transition to the continuum described by fixed-order QCD have been obtained in~\citere{Bach:2017ggt}.

Above threshold, several predictions at fixed order in QCD have been
provided for the on-shell production of a top--antitop pair, \emph{i.e.}\ $\Pe^+\Pe^-\to\Pt\bar\Pt$, reaching next-to-next-to-next-to-leading-order (N$^3$LO) accuracy for the inclusive cross section~\cite{Hoang:2008qy,Kiyo:2009gb} and \sloppy next-to-next-to-leading-order (NNLO) accuracy at the differential level~\cite{Gao:2014nva,Gao:2014eea,Chen:2016zbz,Bernreuther:2023ulo}.
For the off-shell top--antitop pair production with leptonic decays, \sloppy
\emph{i.e.}\ $\eettll$, which is well-defined both below and above
threshold, several next-to-leading-order (NLO) QCD predictions have been provided~\cite{Guo:2008clc,Liebler:2015ipp,ChokoufeNejad:2016qux}.
Regarding electroweak (EW) corrections, NLO accuracy for the inclusive
cross section has been achieved long
ago~\cite{Fujimoto:1987hu,Beenakker:1991ca,Fleischer:2003kk} and
later supplemented with $\order{\alpha^2}$ ISR effects \cite{NhiMUQuach:2017lrx}.
Recently, the QED ISR effects at NLL in collinear factorisation have been
matched to NLO EW corrections for on-shell production \cite{Bertone:2022ktl}.

It is worth emphasising that for off-shell predictions the fully leptonic final state has been usually considered in the literature,
with the exception of some sensitivity and background studies relying on LO off-shell simulations
in the lepton-plus-jets channel \cite{Amjad:2013tlv,Fuster:2014hfw,Amjad:2015mma,Bernreuther:2017cyi}.

In particular, the NLO QCD corrections for the semi-leptonic final state, \emph{i.e.}\ $\eettsl$ are still unknown.\footnote{For the LHC, the semi-leptonic final state has been computed by some of us few years ago~\cite{Denner:2017kzu}.}
The lepton+jets channel has the advantage to possess a larger cross section owing to the larger W-decay branching ratio.
It also allows to fully reconstruct the momenta of the top quarks.

In the present work, we fill this gap by computing for the first time NLO QCD corrections for the process $\eettsl$.
In particular, we discuss phenomenological results in the case where all final-state particles are well separated, which corresponds to a so-called \emph{resolved} topology as opposed to the case where light jets are allowed to be clustered in a large-radius b~jet (\emph{boosted} topology).
We provide cross sections and differential distributions for different CM energies.

A further advancement of this calculation concerns the implementation
of the FKS subtraction scheme~\cite{Frixione:1995ms} in the Monte
Carlo integration code \mocanlo. Among others, the present calculation served to
validate the implementation of the FKS subtraction terms for
processes with only final-state soft and collinear singularities.

This article is organised as follows. In \refse{sec:setup},
the process under investigation is presented \sloppy (\refse{sec:process}), the
input parameters and event selections are listed \sloppy (\refse{subsec:setup}), and
several remarks are provided regarding our implementation \sloppy (\refse{sec:implem}).  Section
\ref{sec:results} discusses numerical results for the fiducial cross
section and differential distributions.
Finally, in \refse{sec:con} the main results obtained are summarised.

\section{Calculation details}\label{sec:setup}
\subsection{Definition of the process}\label{sec:process}
In the present work, we consider the production of a top--antitop pair in $\Pe^+\Pe^-$ collisions in the semi-leptonic decay channel,
\beq\label{eq:resolvedprocess}
\eettsl\,,
\eeq
at NLO QCD accuracy.
All final-state particles (quarks and leptons) are considered massless, and no quark mixing is taken into account (unit CKM matrix).
With such a choice, the NLO corrections of order $\mc{O}(\alphas\alpha^6)$ are genuine QCD corrections
to the leading-order (LO) EW [$\mc{O}(\alpha^6)$] cross section, as the EW corrections to the LO interference [$\mc{O}(\alphas\alpha^5)$] vanish thanks to colour algebra.
The real corrections are made of all possible gluon emissions from any of the coloured particles.
The virtual corrections consist in the interference of Born amplitudes with one-loop ones, which are obtained by the insertion of a gluon in the tree-level matrix elements.

\begin{figure*}
  \centering
  \subfigure[\label{fig:diagA}]{\includegraphics[width=0.29\textwidth]{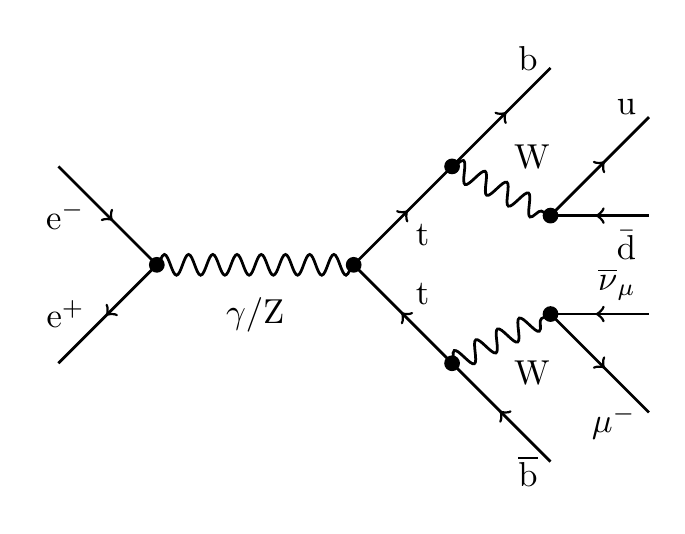}}
  \subfigure[\label{fig:diagB}]{\includegraphics[width=0.29\textwidth]{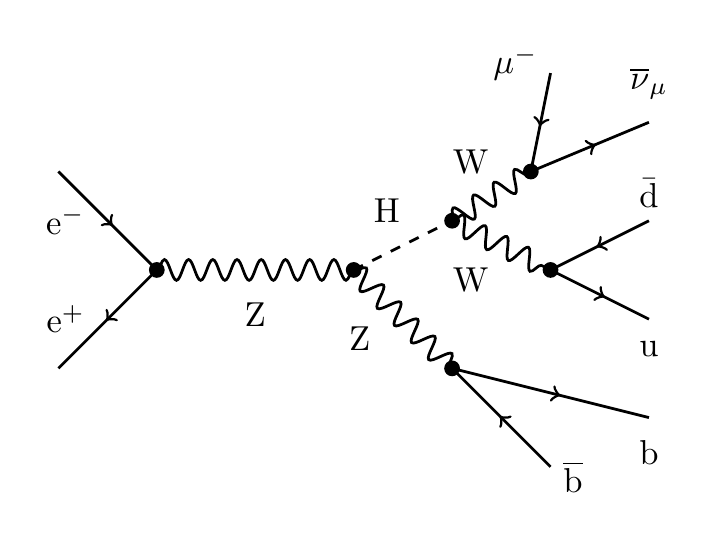}}
  \subfigure[\label{fig:diagC}]{\includegraphics[width=0.29\textwidth]{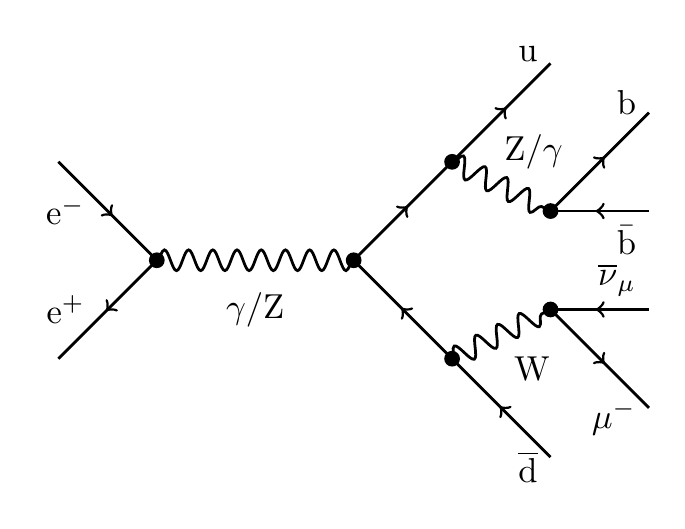}}
  \subfigure[\label{fig:diagD}]{\includegraphics[width=0.29\textwidth]{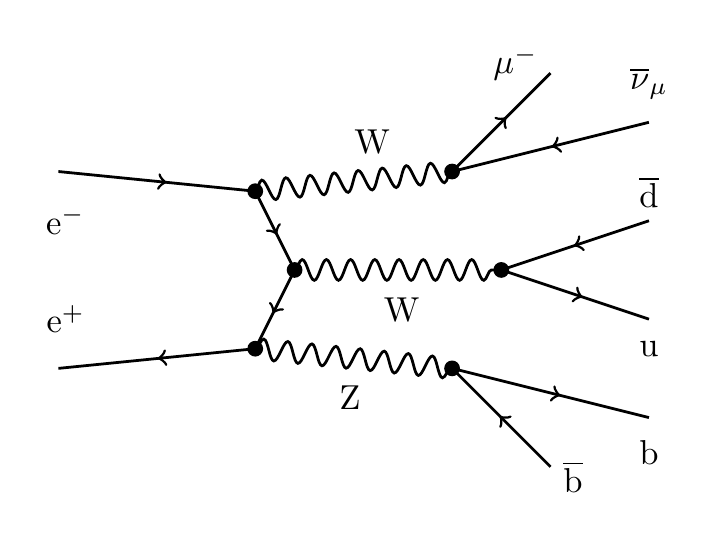}}
  \subfigure[\label{fig:diagE}]{\includegraphics[width=0.29\textwidth]{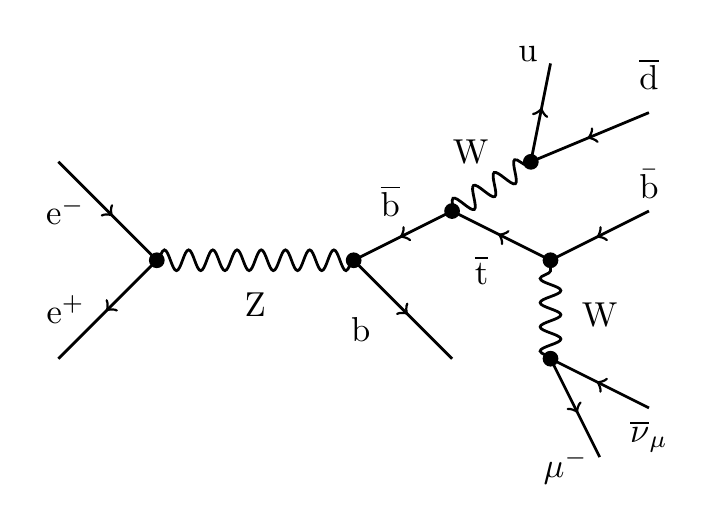}}
  \caption{Sample tree-level Feynman diagrams for the process $\eettsl$.}\label{fig:diag}
\end{figure*}

As illustrated in \reffi{fig:diag} for the leading order, all possible non-resonant and off-shell contributions are accounted for.
In the top row, on the left-hand side, the typical production of a pair of top quarks and their semi-leptonic decay is depicted.
The middle diagram shows a Higgs-strahlung type contribution where the Higgs boson decays into a pair of W~bosons and the Z~boson into a bottom--antibottom pair.
The diagram on  the right-hand side shows a contribution to the
same final state that does not involve any resonant top or antitop
quark or Higgs boson.
In the second row, a tri-boson (left) and a single-top (right) contribution are shown.
We do not consider initial-state-radiation (ISR) and beam-strahlung effects of QED type as we restrict ourselves to QCD corrections.

\subsection{Input parameters and kinematic selections}\label{subsec:setup}

The computation is carried out in the five-flavour scheme, therefore $m_{\Pb}=0$ is assumed throughout.
The on-shell weak-boson masses and decay widths are fixed as \cite{ParticleDataGroup:2020ssz}
\begin{alignat}{2}\label{eq:ewmasses}
    \MWOS &= 80.379 \GeV,&\qquad \GWOS &= 2.085\GeV, \nnb\\
    \MZOS &= 91.1876 \GeV,&\qquad \GZOS &= 2.4952\GeV,
\end{alignat}
and then converted into the pole values \cite{Bardin:1988xt}.
The Higgs-boson and top-quark pole masses are chosen as  \cite{ParticleDataGroup:2020ssz}
\begin{alignat}{2}\label{eq:othermasses}
    \MH &= 125 \GeV,&\qquad \GH &= 4.07 \times 10^{-3}\GeV, \nnb\\
    \Mt &= 173 \GeV,&\qquad \Gt &= 1.3448\GeV.
\end{alignat}
While the Higgs-boson width is taken
from \citere{Heinemeyer:2013tqa}, the numerical value of the top-quark
width is obtained by applying relative QCD corrections from \citere{Basso:2015gca} to the LO top-quark width computed following \citere{Jezabek:1988iv}.
All unstable particles are treated within the complex-mass scheme \cite{Denner:1999gp,Denner:2005fg,Denner:2006ic,Denner:2019vbn}.

The EW coupling constant $\alpha$ is computed within the $G_\mu$ scheme \cite{Denner:2000bj} with the Fermi constant set to 
\beq
\GF = 1.16638\cdot10^{-5} \GeV^{-2}\,.
\eeq

The running of the strong coupling $\alphas$ is carried out at two
loops using the \recola program~\cite{Actis:2016mpe}, assuming  $\as(\MZ)=0.118$.
Finally, the renormalisation scale is set to  $\mu_{\rR}=\Mt$, and
the scale uncertainty is obtained by varying $\mu_{\rR}$ by a factor 2 up and down.

In the following, we consider $\Pe^+\Pe^-$ collisions at several CM energies.
In addition to a scan of the integrated cross sections between $300\GeV$ and
$2\TeV$, shown in \refse{sec:totxs}, we focus on two particular CM energies.
Specifically, in \refse{sec:diff365} we provide differential results for $365\GeV$, \ie the
highest collision energy envisioned for the FCC-ee~\cite{FCC:2018evy}.
A similar energy is planned for the first operating scenario of CLIC
\cite{CLIC:2016zwp,CLICdp:2018esa}, targeting the production of $\Pt\bar{\Pt}$
pairs above threshold.
In \refse{sec:diff1500}, we show differential results for $1.5\TeV$,
\ie the second operating stage of CLIC~\cite{Dannheim:2019rcr}.

In our setup, the jet-clustering is carried out with the $k_{\rm T}$ algorithm \cite{Catani:1993hr} and a resolution radius $R=0.4$ \cite{Boronat:2016tgd,CLICdp:2018esa}.
Note that a generalised version of the $k_{\rm T}$ algorithm has been used in the $\Pt\bar{\Pt}$ study in the fully-leptonic decay channel \cite{ChokoufeNejad:2016qux}.
The clustering algorithm is applied on partons (quarks and gluons) with a minimum angle of $0.7721^{\circ}$, which corresponds to a rapidity of 5.
We use the following selection cuts, which are inspired by event selections applied in CLIC and FCC-ee studies~\cite{Boronat:2016tgd,CLICdp:2018esa,Dannheim:2019rcr}.
In our calculation, the events are required to have:
  \begin{itemize}
  \item[$\bullet$] a minimum missing transverse-momentum $\pt{\rm miss}>20\GeV$, which is defined as the transverse momentum of the neutrino;
  \item[$\bullet$] a minimum transverse momentum $p_{\rT}>20\GeV$ for
    the antimuon, the light jets, and the two b-tagged jets;
  \item[$\bullet$] an angular acceptance of $10^{\circ}<\theta<170^{\circ}$ for the antimuon, the light jets, and the two b-tagged jets;
  \item[$\bullet$] a minimum rapidity--azimuthal-angle distance between
    the antimuon and the jets, $\Delta R_{\rm \ell j},\Delta R_{\rm \ell j_{\rm b}}  > 0.4$;
  \item[$\bullet$] an invariant-mass cut on the system formed by the two
    hardest visible
    light jets, the charged lepton, and the neutrino of $M_{\Pj\Pj\mu^+\nu_\mu}>130\GeV$.
  \end{itemize}
We require at least two light jets that fulfil all requirements on
the transverse momentum, the angular acceptance, and the
rapidity--azimuthal-angle distance to leptons (visible jets). Out of these jets, we
select the two hardest ones (according to transverse momentum) for the
distributions shown below.
The condition on the invariant mass $M_{\Pj\Pj\mu^+\nu_\mu}$ selects a kinematic region that excludes the
Higgs-boson decay into two leptons and two jets.

The two b~jets present in the final state can be
associated with the leptonically decaying top quark ($\bj^{\Pt_{\rm lep}}$)
and the hadronically decaying antitop quark ($\bj^{\Pt_{\rm had}}$).
This is achieved by finding the maximum of a likelihood function that
is the product of two Breit--Wigner distributions (of the top and antitop quark), as done in \citere{Denner:2020orv}.
The likelihood function mimics the top- and antitop-quark propagators,
assuming three-body decays after recombination (both at LO and at NLO QCD),
and reads
\begin{align}
\label{eq:likelihood}
 \mathcal{L}_{ij} =& \frac{1}{\left(p^2_{\mu^+\nu_\mu\Pj_{\Pb,i}} -
     m_\Pt^2\right)^2+\left(m_\Pt \Gamma_\Pt\right)^2} \; \nnb\\
 &\times\,\frac{1}{\left(p^2_{\Pj\Pj\Pj_{\Pb,j}} - m_\Pt^2\right)^2+\left(m_\Pt \Gamma_\Pt\right)^2}\, ,
\end{align}
with $p_{a b c} = p_{a} + p_{b} + p_{c}$.  The combination of bottom
jets $\{\Pj_{\Pb,i}, \Pj_{\Pb,j}\}$ that maximises $\mathcal{L}_{ij}$
defines the two bottom jets originating from the leptonic and hadronic
top quarks.  Note that in Eq.~\eqref{eq:likelihood} all possible
combinations of light jets and b~jets are considered. This includes
also light jets with a minimum angle of $0.7721^{\circ}$
that do not fulfil the
transverse-momentum, angular, and rapidity--azimuthal-angle distance
requirements for visible jets.

Note further that the neutrino momenta are extracted from Monte Carlo
truth, assuming that the hard-scattering CM energy is exactly the one
of the $\Pe^+\Pe^-$ collision, \ie neglecting ISR and beam-strahlung
effects.

\subsection{Implementation and validation}\label{sec:implem}

To carry out the present calculation, we have employed the Monte Carlo program \mocanlo.
In the past, \mocanlo has been successfully used for several
\sloppy top-associated computations at NLO QCD and/or EW accuracy at hadron colliders~\cite{Denner:2015yca,Denner:2016jyo,Denner:2016wet,Denner:2017kzu,Denner:2020orv,Denner:2020hgg,Denner:2021hqi,Denner:2022fhu}.
The present work is the first application of \mocanlo to a lepton-collider process.
The program uses phase-space mappings similar to those of \citeres{Berends:1994pv,Denner:1999gp,Dittmaier:2002ap} and has shown to be particularly efficient for NLO calculations for high-multiplicity processes (up to $2\to8$).
The tree-level and one-loop matrix elements are obtained from
\recola~\cite{Actis:2012qn,Actis:2016mpe} using the integral library
\collier~\cite{Denner:2016kdg}.
For the subtraction of infrared divergences, the original code relies on the Catani--Seymour subtraction scheme~\cite{Catani:1996vz,Dittmaier:1999mb,Dittmaier:2008md}.
For the present calculation, we have implemented the FKS
scheme~\cite{Frixione:1995ms} following closely \citeres{Frixione:2007vw,Frederix:2009yq}.

\begin{figure*}
  \centering
  \subfigure[\label{fig:comp_cos}]{\includegraphics[width=0.42\textwidth]{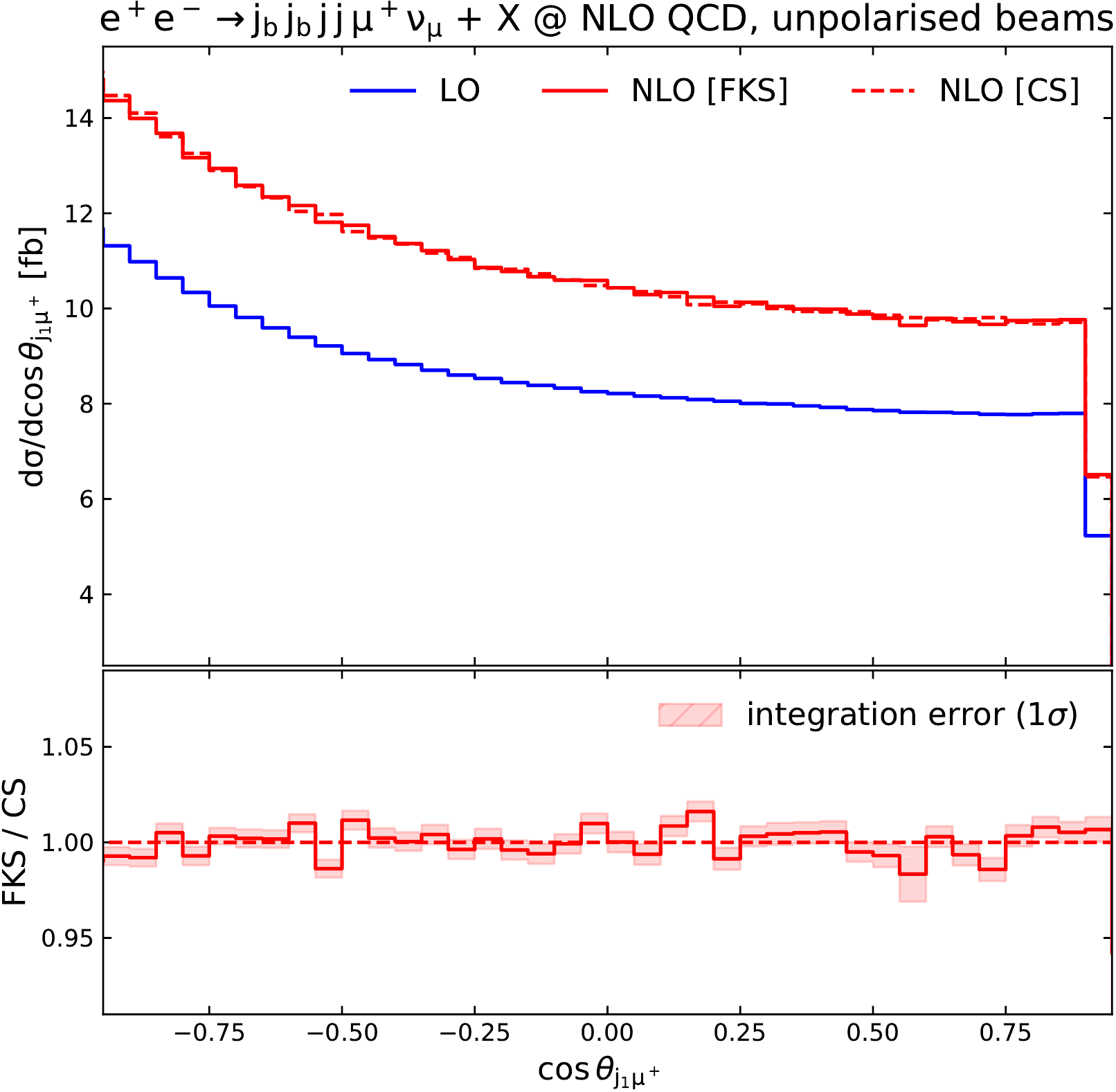}}\qquad\qquad
  \subfigure[\label{fig:comp_pt}]{\includegraphics[width=0.42\textwidth]{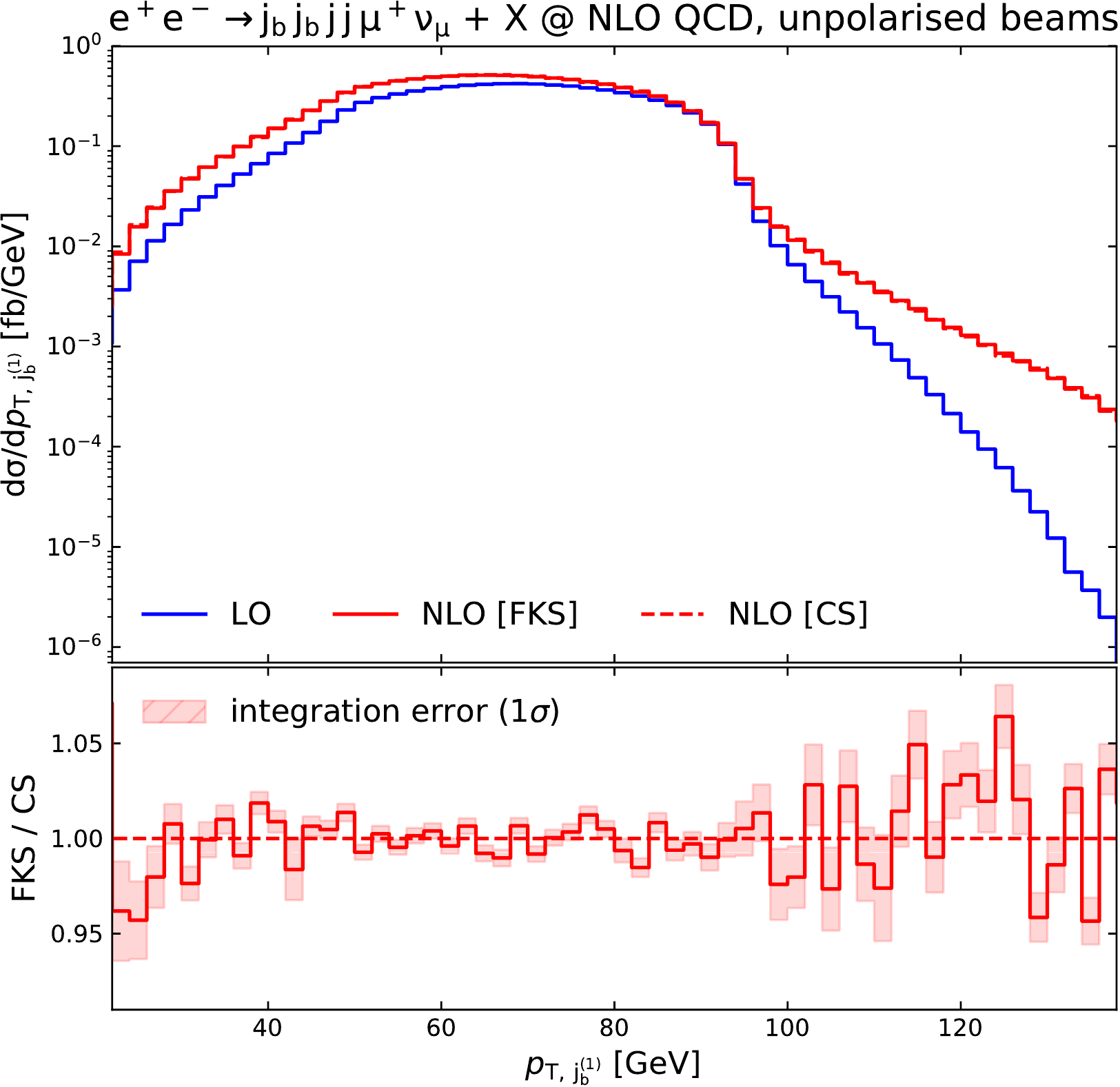}}
        \caption{
          Differential distributions in the cosine of the angular separation between the hardest light jet
          and the antimuon (left) and in the transverse momentum of the hardest b jet (right).
          The NLO QCD distributions obtained with the Catani--Seymour (dubbed CS) and FKS subtraction schemes
          are plotted in the top panel besides the LO results.  
          In the bottom panel the ratio between the FKS and CS results (solid line)
          and the corresponding Monte~Carlo integration error (shaded band) are shown.          
        }.\label{fig:comparison}%
\end{figure*}

To validate our implementation of the  FKS scheme, we have compared
our results against those obtained with the well-tested
Catani--Seymour scheme for several NLO QCD calculations at lepton
colliders including di-jet production, di-boson production in the
semi-leptonic channel, off-shell top--antitop production in the fully
leptonic channel, and the process considered in this work.
In all cases, we have found perfect agreement within the Monte Carlo
uncertainty at the level of both fiducial cross sections and differential
distributions.
For the $\sqrt{s}=365\GeV$ setup considered in this article,
the fiducial cross section obtained at NLO QCD with the two subtraction schemes
reads, 
\begin{align}
\sigma_{\rm NLO}^{\rm (FKS)}& =  21.419(14)^{+2.2\%}_{-1.8\%}\,\fb\,,\\
\sigma_{\rm NLO}^{\rm (CS)} & =  21.427(12)^{+2.2\%}_{-1.8\%}\,\fb\,,
\end{align}
respectively, showing perfect agreement. Agreement has also been found at the
differential level, where the numerical differences between the NLO distributions
obtained with the two schemes are well within integration uncertainties bin by bin.
We provide in \reffi{fig:comparison} a comparison of the differential results obtained with the two subtraction schemes for two observables, namely an angular one and a transverse-momentum one.
%
The integration errors displayed in the lower inset of the plots are the combined integration errors of both NLO calculations.
We observe that the bin-wise agreement is within the integration uncertainties over the whole spectrum.
An analogous picture has been found for all other observables that we have computed.

In addition to the comparison against the dipole formalism, our
implementation of the FKS subtraction scheme has allowed for further tests:
\begin{itemize}
\item[$\bullet$] The cancellation of infrared poles in the $n$-body contribution to the NLO QCD cross section has been verified by evaluating the virtual contribution $\mc V$ and the integrated FKS counterterm $\mc I$ at different values of the infrared scale $\mu_{\rm IR}$ (from $10^{-8}\GeV$ to $10^{8}\GeV$) and checking that the sum $\mc V+\mc I$ is independent of the $\mu_{\rm IR}$ choice.
This has been carried out for a large number of phase-space points
finding agreement up to 12 digits.
\item[$\bullet$] The cancellation of phase-space singularities between the
  real matrix element and the FKS subtraction counterterm has been verified by
  constructing real-phase-space points that approach the soft, collinear,
  and soft--collinear regions by means of a rescaling of the radiation variables.
\item[$\bullet$] The FKS-subtraction parameters $\xi_{\rm c}$ and $\delta$
  \cite{Frixione:1995ms,Frixione:2007vw,Frederix:2009yq} which define the integration 
  boundaries for the soft and collinear regimes, have been varied,
  confirming that the sum of the subtraction counterterm and its integrated counterpart is
  independent of them. Selected results regarding this subtraction test are shown in
  Table~\ref{tab:tech}, where the reader can observe a rather strong impact of the soft
  parameter $\xi_{\rm c}$ on the size and sign of subtracted real and virtual contributions, compared to a milder effect of the collinear one $\delta$.
  \begin{table*}
    \begin{center}
      \begin{tabular}{cc|ccc}
        $\xi_{\rm c} $ & $\delta $ & $\mc R^{\rm subtr} $ &  $ \mc V^{\rm \,subtr} $ & $\mc R^{\rm subtr} + \mc V^{\rm\, subtr} $  \rule[-1ex]{0ex}{2.5ex}\\
       \hline\rule{0ex}{2.7ex}%
        0.01 & 0.01 & $37.497(4)$ & $-32.94(1)$ & $4.55(2)$\\
        0.4  & 0.01 & $-15.02(2) $ & $19.51(7) $ & $4.49(7) $ \\
        0.01 & 0.4  & $6.14(2) $ & $-1.620(9) $ & $4.53(2) $ \\
        0.4  & 0.4  & $-4.38(2) $ & $8.86(5) $ & $4.49(5) $        
      \end{tabular}
    \end{center}
    \caption{Comparison of subtracted virtual ($\mc V^{\rm \,subtr} $) and real ($\mc R^{\rm subtr} $)
      contributions to the NLO QCD correction to $\eettsl$
      for different choices of FKS-subtraction soft ($\xi_{\rm c}$) and collinear ($\delta$) parameters.\label{tab:tech}}
  \end{table*}
\item[$\bullet$] The evaluation of FKS sector functions and their sum rules have been checked
  in the subtracted-real contribution by means of a variation of the exponents $a,\,b$ that
  enter the sector functions, which are defined in Eq.~(5.11) of \citere{Frederix:2009yq}. 
  For a fixed choice of the FKS parameters ($\xi_{\rm c}=\delta=0.01$) we have calculated the
  subtracted-real contributions for different choices of such exponents, \eg
  \begin{align}
  a,b=1:\quad\sigma^{\rm subtr}_{\rm real}\,&=\, 37.423(40)\fb \,,\nnb\\
  a,b=4:\quad\sigma^{\rm subtr}_{\rm real}\,&=\, 37.451(41) \fb \,,
  \end{align}
  finding perfect agreement within integration errors.
\end{itemize}
The default values used for the results presented here are
$\xi_{\rm c}=\delta=0.01$ and $a=b=1$.

\section{Results}\label{sec:results}

\subsection{Fiducial cross sections}\label{sec:totxs}
\begin{figure*}
  \hfill
  \vspace*{-6cm}\includegraphics[width=0.41\textwidth]{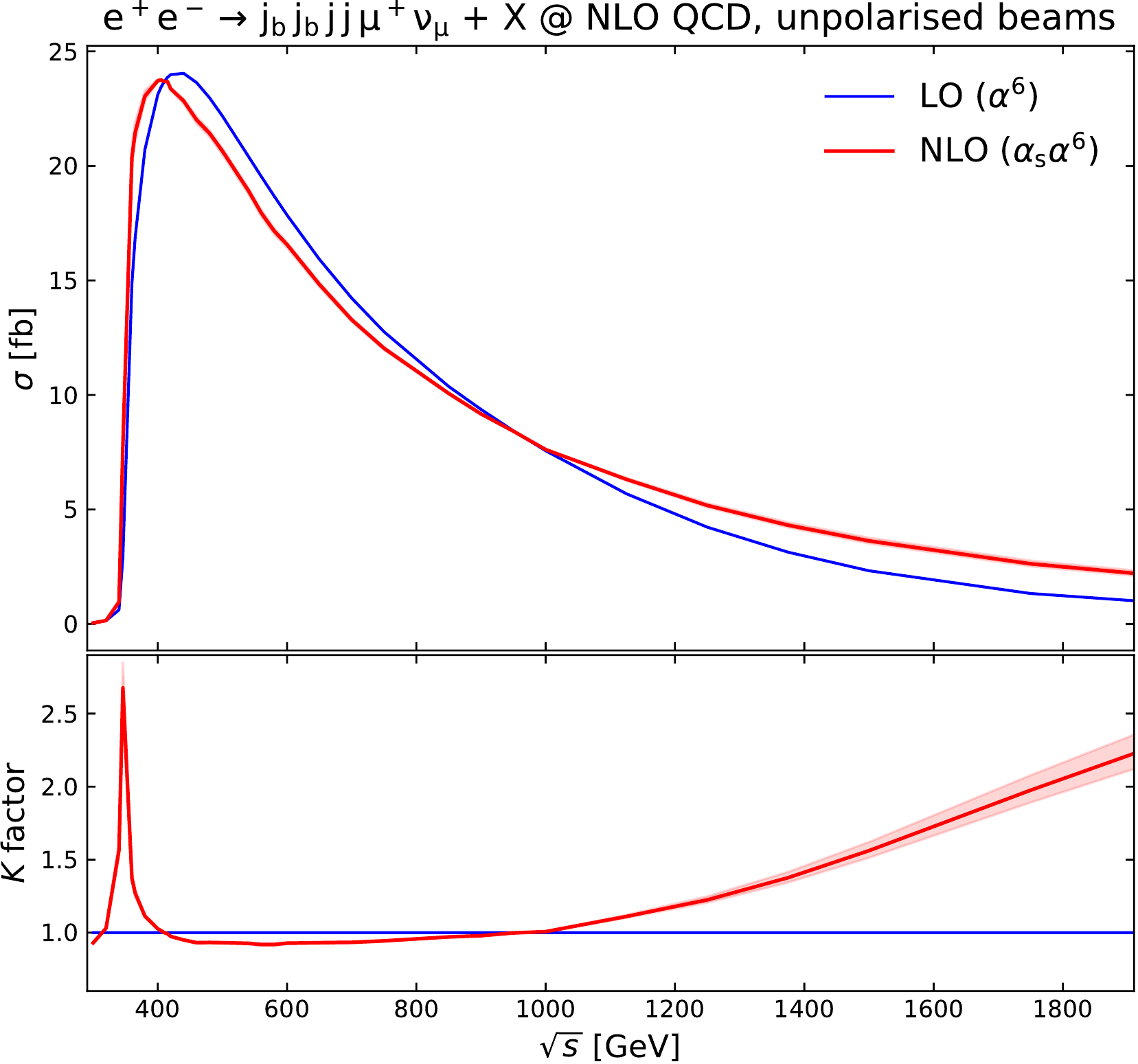}\\
  \begin{tabular}{c|c|c|c}
    $\sqrt{s}$ [$\GeV$] & $\sigma_{\rm LO}$ [$\fb$] & $\sigma_{\rm NLO
      \, QCD}$ [$\fb$] & $\delta_{\rm NLO \, QCD} [\%]$
    \rule[-1.5ex]{0ex}{2.5ex} \\
    \hline\rule{0ex}{3ex}%
 320  &  0.14461(5) &   0.1490(8) $^{ +  0.3\% }_{  -0.3\% }$  &   $  3.03  $  \\[0.1cm]
 340  &  0.6153(2)  &   0.963(3)$  ^{ +  3.8\% }_{  -3.1\% }$  &   $  56.6 $  \\[0.1cm]
 346  &  2.9127(8)  &   7.795(6)$  ^{ +  6.6\% }_{  -5.4\% }$  &   $  167.6$  \\[0.1cm]
 365  &  16.877(4)  &   21.42(2)$  ^{ +  2.2\% }_{  -1.8\% }$  &   $  26.9 $  \\[0.1cm]
 405  &  23.437(7)  &   23.75(4)$  ^{ +  0.1\% }_{  -0.1\% }$  &   $  1.34  $  \\[0.1cm]
 440  &  24.040(8)  &   22.84(9)$  ^{ +  0.5\% }_{  -0.6\% }$  &   $  -5.01 $  \\[0.1cm]
 560  &  19.542(7)  &   17.95(4)$  ^{ +  0.8\% }_{  -0.9\% }$  &   $ -8.15  $  \\[0.1cm]
1125  &  5.683(1)   &   6.31(3) $  ^{ +  1.1\% }_{  -0.9\% }$  &   $  11.1 $  \\[0.1cm]
1500  &  2.3235(8)  &   3.627(9)$  ^{ +  3.8\% }_{  -3.1\% }$  &   $  56.1 $\vspace*{0.8cm}
    \end{tabular}
  \caption{
    Fiducial cross sections for $\eettsl$ at LO and NLO QCD at various CM energies $\sqrt{s}$.
    Numerical values are shown in the table (left), where the digit in parentheses indicates the Monte Carlo statistical error,
    while the sub- and super-scripts in per cent indicate the renormalisation-scale uncertainties calculated
    with three-point scale variations. The integrated cross section is shown in the figure (right)
    at LO (blue) and NLO QCD (red) as a function of the CM energy. The red-shaded band is obtained
    by means of three-point renormalisation-scale variations.
    }\label{tab:XsectionQCD}
\end{figure*}

In this section, we report results for the fiducial cross section
in the setup defined in \refse{sec:setup}.  In \refta{tab:XsectionQCD}, we provide
the results at LO and NLO QCD accuracy for several choices of the CM energy, including
regimes below and above the $\Pt\bar{\Pt}$ threshold. 
Since $\eettsl$ is a purely EW process at LO, there is no scale dependence at this order.
At NLO, the QCD uncertainty comes from three-point variations of the renormalisation scale,
$\mu_{\rm R}/\mu^{(0)}_{\rm R}=1/2,\,1,\,2$.

The NLO QCD corrections are strongly dependent on the CM energy as can be seen in \reffi{tab:XsectionQCD},
where the fiducial cross section is provided as a function of the CM energy
in the range from $300\GeV$ to $2\TeV$.
The largest QCD corrections are observed slightly below the top--antitop threshold.
In this regime, the presence of the Coulomb singularity renders the NLO QCD corrections divergent
for on-shell top quarks \cite{Melnikov:1993np,Guo:2008clc}, while the inclusion
of decay effects makes them finite though still very large, reaching almost $170\%$ of the LO cross section.
In the case of the fully leptonic top-quark decays, a similar behaviour is observed
at and around threshold~\cite{Guo:2008clc,ChokoufeNejad:2016qux}.
Above threshold the corrections turn negative.
For the semi-leptonic process, the NLO QCD corrections become positive for energies above $1\TeV$.
For example at $1.5\TeV$, the corrections are very large of the order of $60\%$.
This is in contrast with the fully leptonic case where the corrections stay negative at high energies.
This difference is due to the specific event selection.
At LO, the cross section is suppressed in the semi-leptonic channel by
the jet clustering (with $R=0.4$) which effectively forbids boosted $\PW$~bosons decaying into two quarks.
At NLO QCD, in the presence of real gluon radiation, this constraint is lifted for sufficiently hard gluons.
This part of the phase space therefore opens up and leads to relatively large corrections.
This effect is specific to the semi-leptonic final state as in the fully leptonic case
there are no cuts preventing boosted $\PW$~bosons.
We have verified this explanation by running the calculation for different jet radii.
Smaller relative QCD corrections are found for
a smaller jet-clustering radius ($R=0.1$, allowing configurations with more
boosted $\PW$~bosons). For instance, at $\sqrt{s}=1\TeV$ we find:
\begin{align}
 &R=0.4\,:\quad \sigma_{\rm LO} \, =  2.3235(8)\,\fb\,,\quad \delta_{\rm QCD}\,=+56.1\%\nnb,\\
 &R=0.1\,:\quad  \sigma_{\rm LO} \, =  4.1524(6)\,\fb\,,\quad \delta_{\rm QCD}\,=+4.8\% .
\end{align}

The scale uncertainty increases from sub-percent to $5-7\%$ when approaching
the threshold ($\sqrt{s}\lesssim 346\GeV$) owing to the large QCD
corrections in this region.
For $400\GeV\lesssim\sqrt{s}\lesssim 1\TeV$,
it decreases down to sub-percent level, while for $\sqrt{s}\gtrsim 1\TeV$ it
increases up to $\mc O(5{-}10\%)$ level driven by large real-radiation corrections.

Finally, we mention that the off-shell calculation embeds $\Pt\bar\Pt$ contributions as well as
irreducible-background contributions that become more and more important
with increasing CM energy,
as observed in ILC sensitivity studies at $500\GeV$
\cite{Amjad:2013tlv,Amjad:2015mma,Bernreuther:2017cyi}.
We have checked numerically that background contributions become indeed relevant.
In particular, we have found that single-top topologies
[\eg $\Pe^+\Pe^-\rightarrow \bar\Pb \PW^- \Pt, \Pb \PW^+ \bar\Pt$, as shown in \reffi{fig:diagE}] are the largest of these contributions.
Tri-boson topologies
[$\Pe^+\Pe^-\rightarrow\PW^+\PW^-\PZ$, as shown in \reffi{fig:diagD}] also contribute but to a lesser extent.
It is therefore interesting to realise that at very high energy the final state under investigation is not only made of top--antitop topologies but also of many others, rendering the reconstruction of top--antitop pairs difficult.

\subsection[Differential distributions at $365\GeV$]
{Differential distributions at \boldmath $365\GeV$}\label{sec:diff365}

In this section, several differential distributions are presented at LO and NLO QCD accuracy.
While the upper panels of the plots contain the absolute predictions, the lower ones show the corresponding $K$~factors.
In the following, when an observable refers to either the leptonically
or the hadronically decaying top quark, their definition follows from the maximisation of the likelihood function in Eq.~\eqref{eq:likelihood}.

\begin{figure*}
  \centering
  \subfigure[\label{fig:pt1}]{\includegraphics[width=0.41\textwidth]{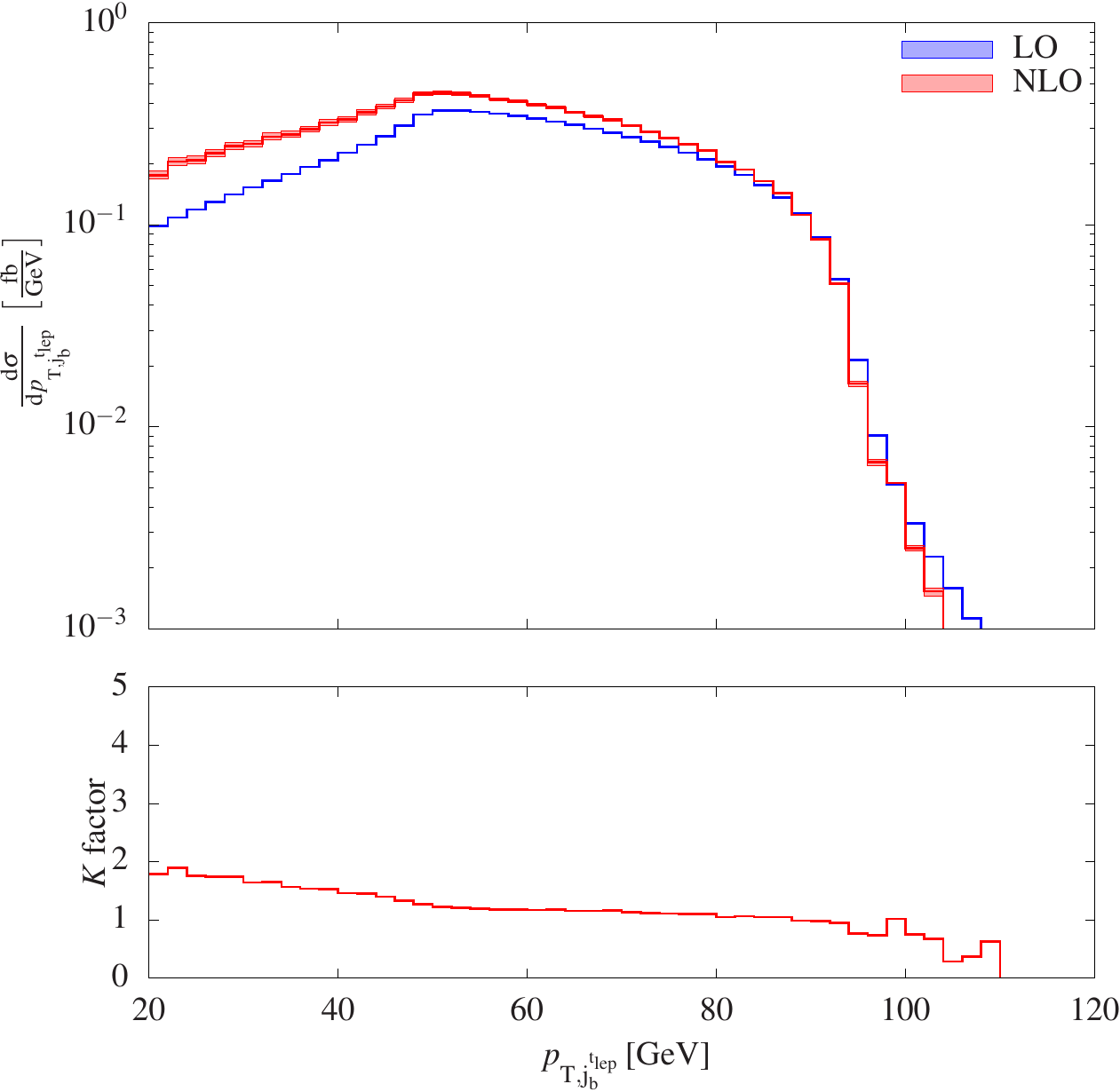}}
  \subfigure[\label{fig:pt2}]{\includegraphics[width=0.41\textwidth]{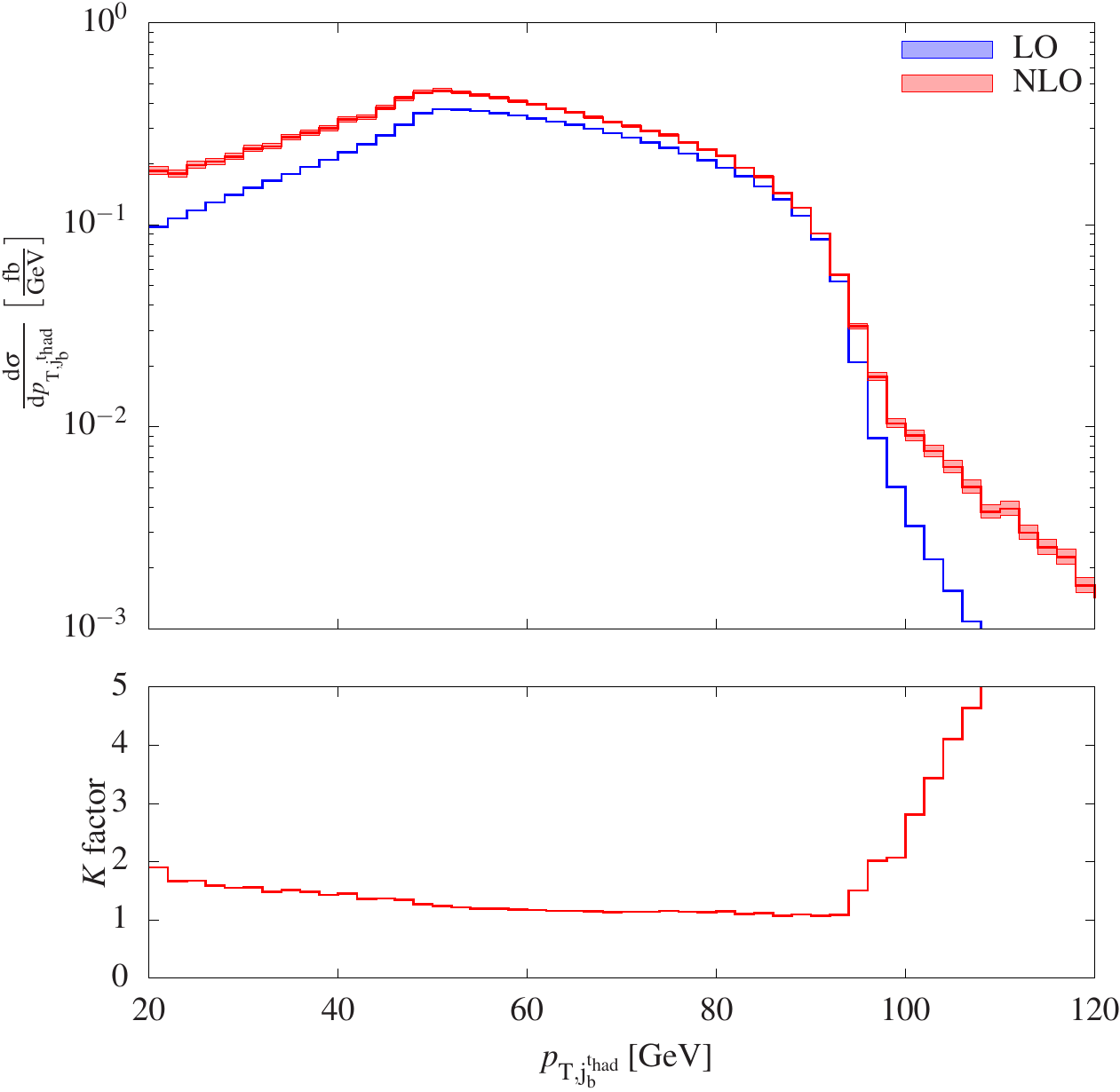}}\\
  \subfigure[\label{fig:pt3}]{\includegraphics[width=0.41\textwidth]{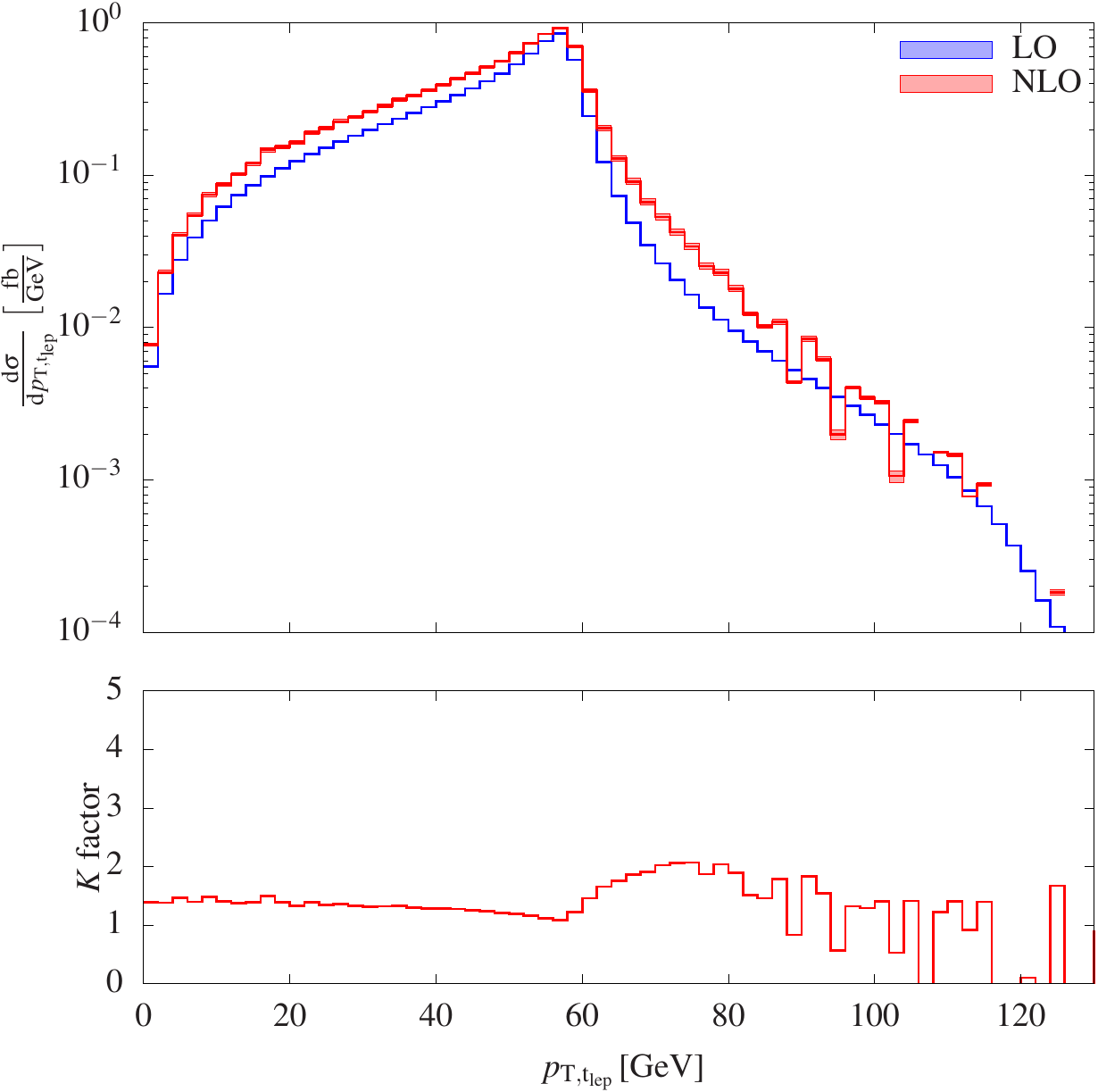}    }
  \subfigure[\label{fig:pt4}]{\includegraphics[width=0.41\textwidth]{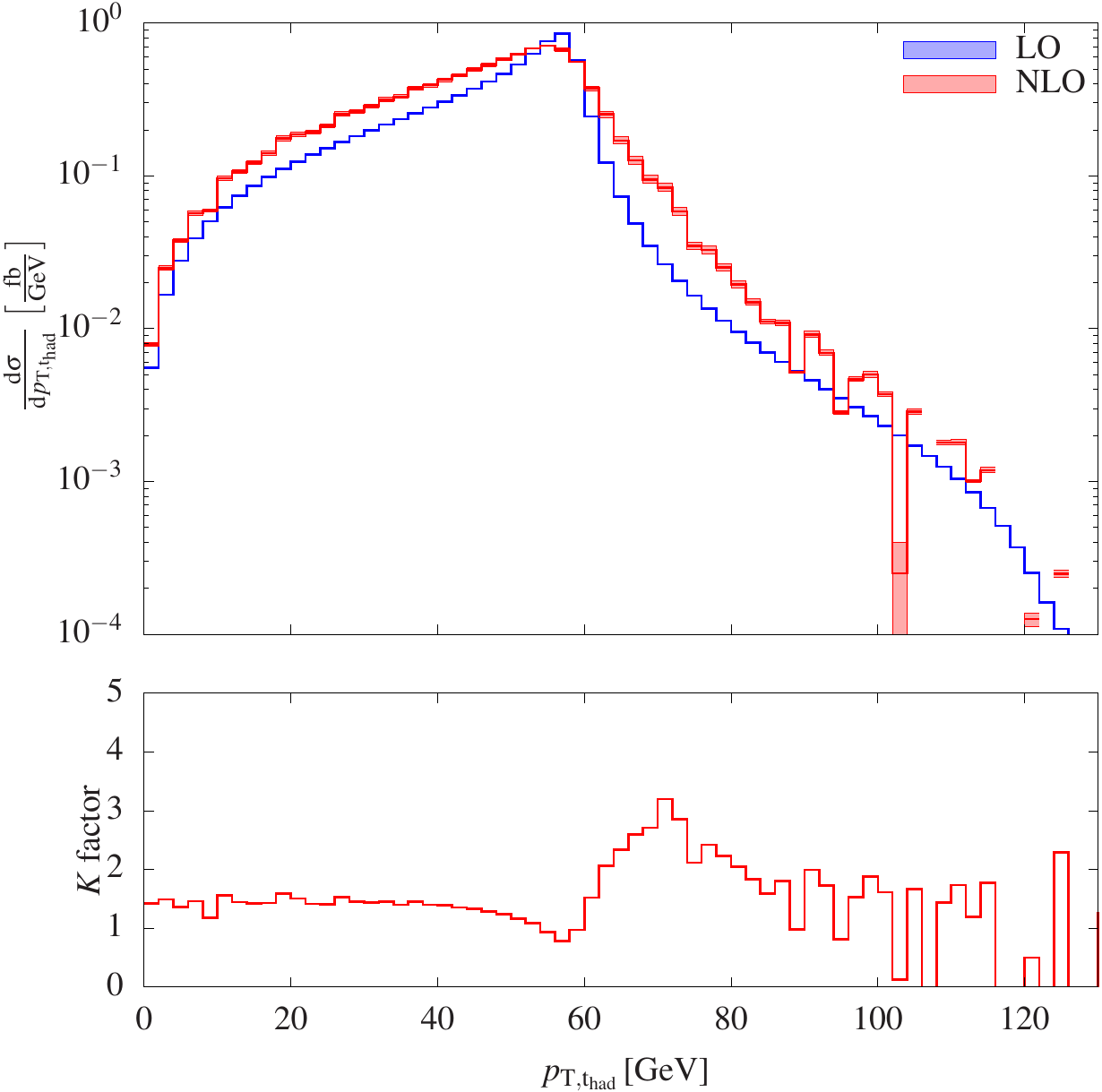}    } 
  \caption{
    Differential distributions for a $365\GeV$ CM energy
    in the transverse momentum of
    the reconstructed bottom quark from the leptonically decaying top
    quark (top left), 
    the reconstructed bottom quark from the hadronically decaying top
    quark (top right),
    the reconstructed leptonically decaying top quark (bottom left), and
    the reconstructed hadronically decaying top quark (bottom
    right). 
    The red-shaded band is obtained by means of three-point renormalisation-scale variations and the lower panel displays the $K$~factor.
  }\label{fig:pT}
\end{figure*}

In \reffi{fig:pT}, several transverse-momentum distributions are shown.
The first two are for the reconstructed bottom quark from the leptonically and hadronically decaying top quarks, respectively.
At LO, both distributions are almost identical with a pronounced drop
around $95\GeV$. This results from a hard cut for production of
on-shell top-quark pairs decaying into on-shell W~bosons that can be evaluated to
\beq
\label{eq:pTjblimit}
p_{\rT,\bj} < \frac{\sqrt{s}}{4}\left(1-\frac{\MW^2}{\Mt^2}\right)
\left[1+\sqrt{1-\frac{4\Mt^2}{s}}\right].
\eeq
For $\sqrt{s}=365\GeV$, this amounts to $p_{\rT,\bj} \lesssim
94.4\GeV$.  At NLO QCD, the picture changes as the hadronic top
quark receives significantly more corrections at high transverse
momenta. This is related to  additional real-radiation jets that are
mis-reconstructed as top-decay jets.
It is worth mentioning that the reconstructed bottom-quark
distributions are very close to those obtained with Monte Carlo
truth.  For what concerns the reconstructed top quarks (leptonic and
hadronic), both at LO and NLO QCD accuracy the two distributions show
very similar qualitative behaviours. On-shell production of
top--antitop quarks is restricted to $p_{\rT,\Pt}<(\sqrt{s}/2)\sqrt{1-4\Mt^2/s}\approx58\GeV$,
leading to a sharp drop of the distribution above this value. In the
off-shell region, the NLO QCD corrections are somewhat larger for the
hadronically decaying top quark. For $p_{\rT,\Pt}\gtrsim127\GeV$ the
recoiling system cannot contain a resonant W~boson anymore explaining
the shoulder near $120\GeV$.
We note that in both cases, around $100\GeV$ and above, the
top-transverse-momentum distributions become numerically unstable.
Besides the lower statistics, this is simply due to the fact that at such energies, the process is not exclusively made of top-antitop topologies as explained in the previous section.

\begin{figure*}
  \centering
  \subfigure[\label{fig:inv1}]{  \includegraphics[width=0.40\textwidth]{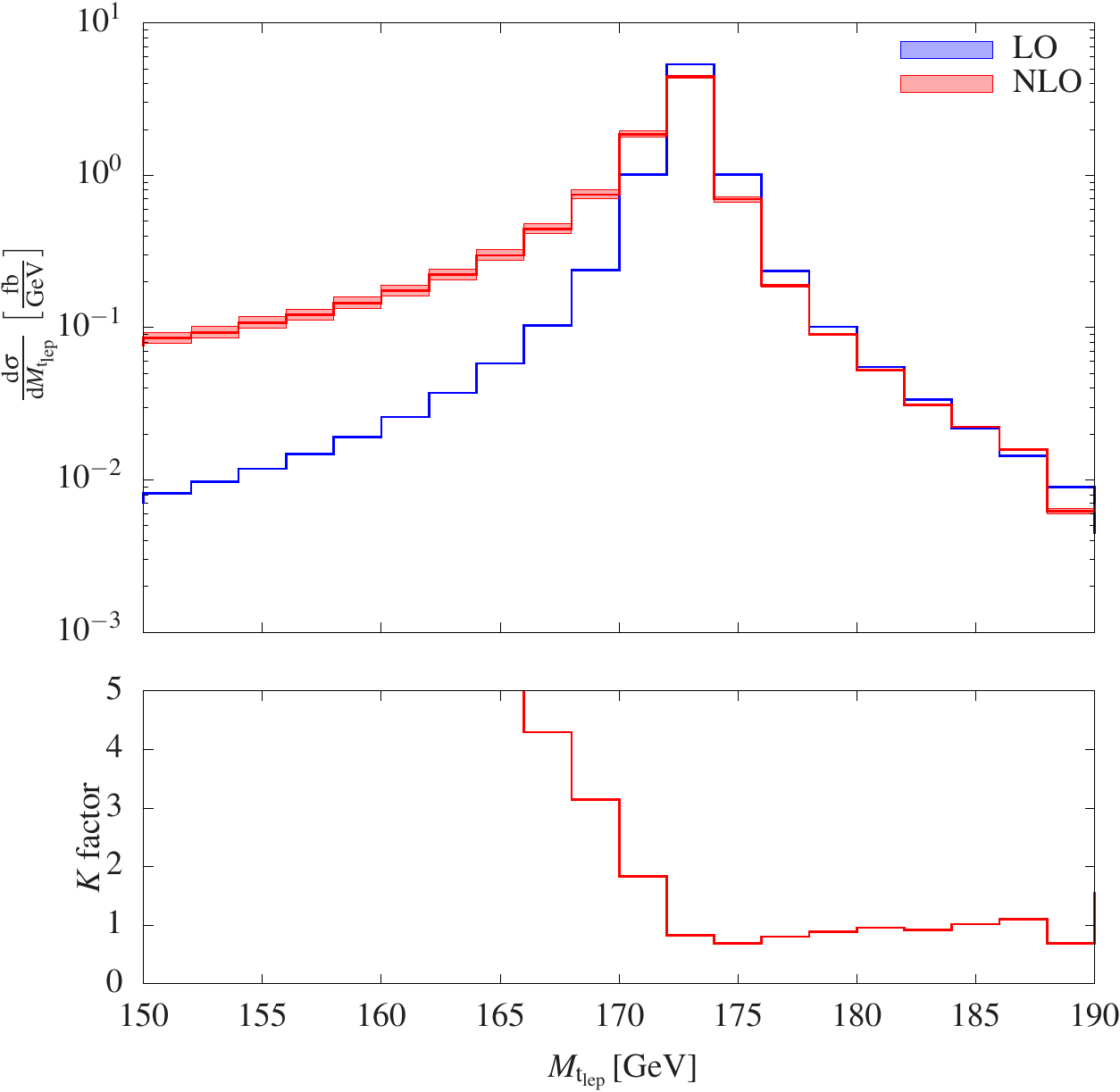}}
  \subfigure[\label{fig:inv2}]{  \includegraphics[width=0.40\textwidth]{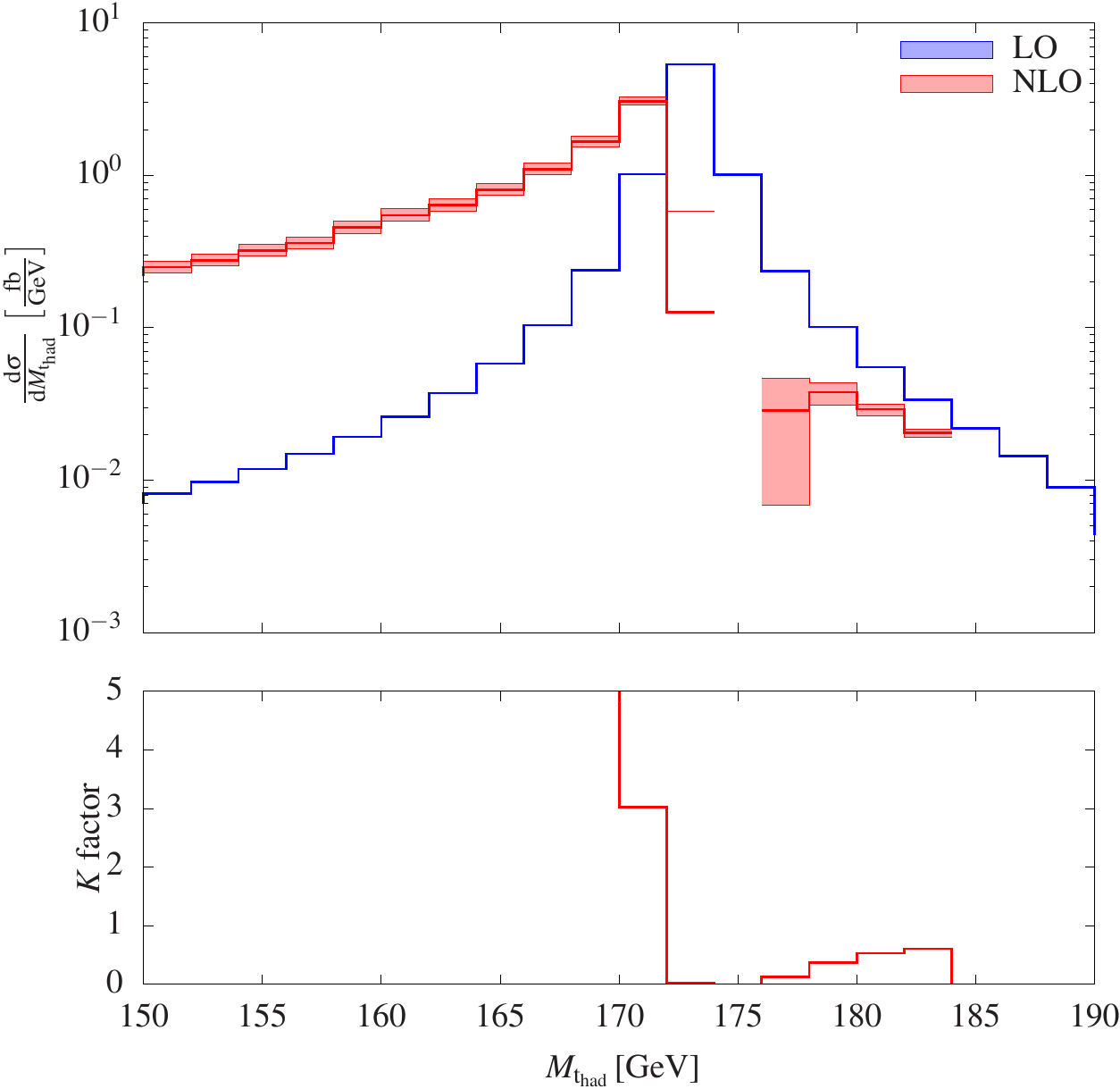}}\\
  \subfigure[\label{fig:inv3}]{  \includegraphics[width=0.40\textwidth]{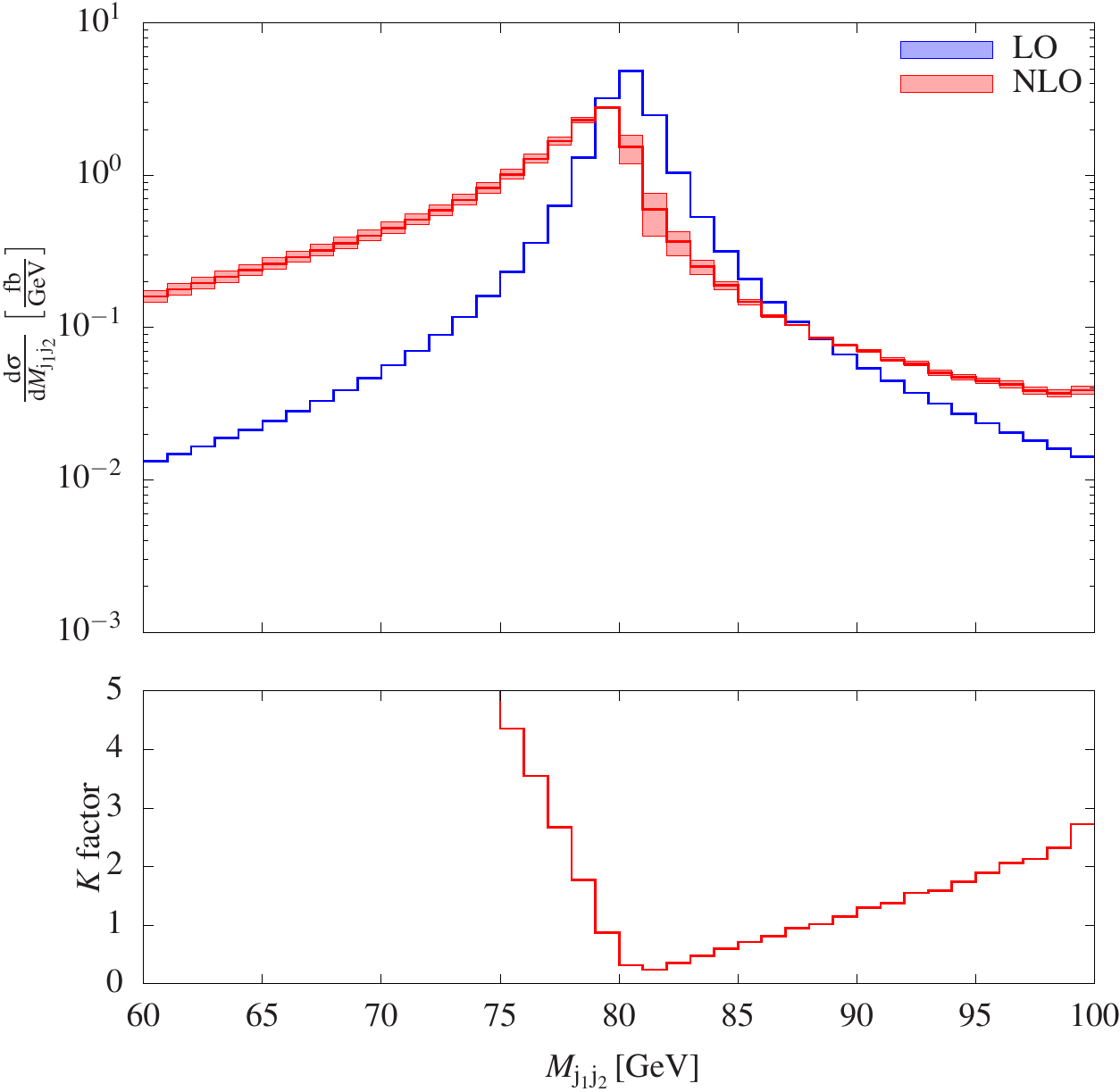}                      }
  \subfigure[\label{fig:inv4}]{  \includegraphics[width=0.40\textwidth]{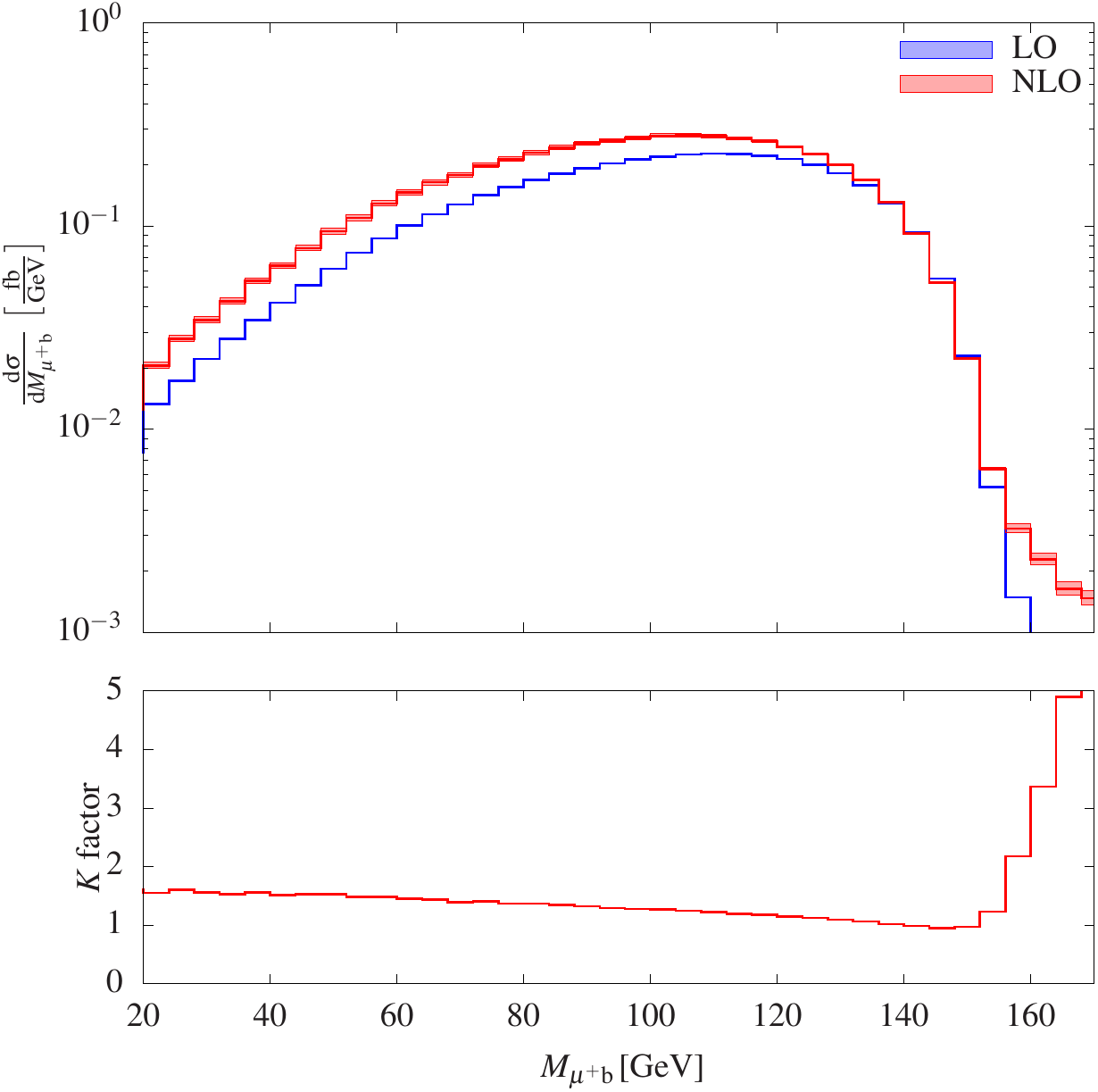}              }\\
  \subfigure[\label{fig:cos1}]{  \includegraphics[width=0.38\textwidth]{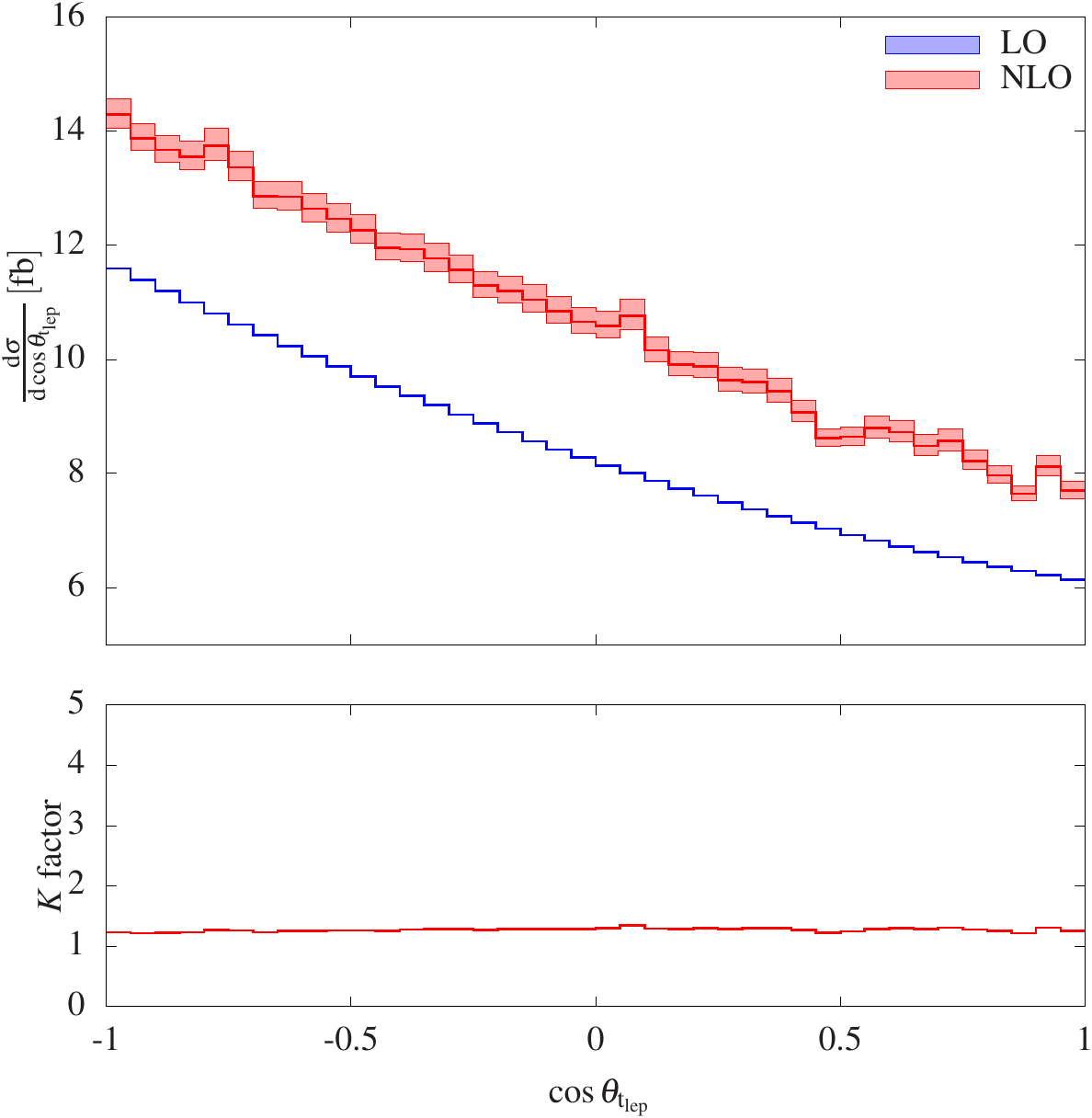}}\qquad
  \subfigure[\label{fig:cos2}]{  \includegraphics[width=0.38\textwidth]{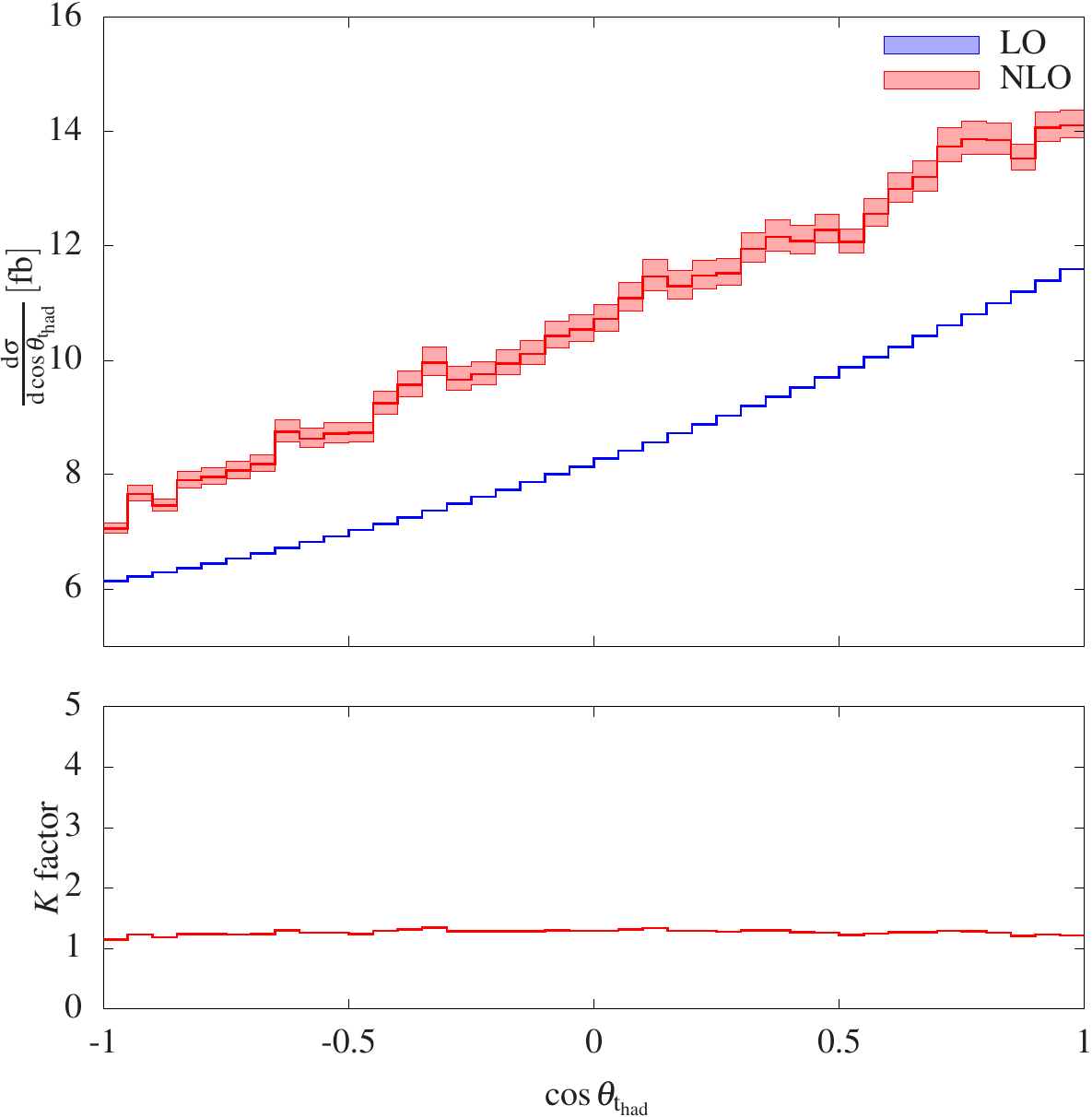}}
  \caption{ 
    Differential distributions for a $365\GeV$ CM energy
    in the invariant masses of
    the reconstructed leptonically decaying top quark (a), 
    the reconstructed hadronically decaying top quark (b),
    the system of the two hardest light jets (c), and
    the bottom quark from the leptonically
    decaying top quark with the antimuon (d), as well as 
    in the cosine of the production angle of the reconstructed leptonically decaying
    top quark (e) and the reconstructed hadronically decaying
    top quark (f).
    Note that for \reffi{fig:inv4}, the Monte Carlo truth momenta are used.
    Same structure as in \reffi{fig:pT}.}\label{fig:inv}
\end{figure*}

In \reffi{fig:inv}, several invariant-mass distributions are displayed.
The first two are for the invariant masses of the leptonically
[\reffi{fig:inv1}] and
hadronically [\reffi{fig:inv2}] decaying top quarks, respectively.
It is interesting to observe that, as for the transverse-momentum distributions,
the LO predictions are essentially identical while at NLO they significantly differ.
This is due to the fact that the hadronically decaying top quark
possesses three partons in the final state as opposed to only one for
the leptonically decaying one, leading to more final-state radiations in the hadronic case.
As a consequence, more events are moved from the resonance
or above to below the resonance owing to final-state radiation that is
not reconstructed with the decay products of the top/antitop quark
forming hence a large radiative tail (see, for instance, \citeres{Denner:2012yc,Denner:2017kzu}).
In the case of the hadronically decaying top quark, the effect is so
large that the NLO cross section becomes negative above the resonance.
Such a behaviour has already been observed for the same final state at
a hadron collider~\cite{Denner:2017kzu} and requires the inclusion of
higher-order corrections for a proper description of this observable.

A radiative tail also appears in the invariant-mass distribution of
the two hardest light jets [\reffi{fig:inv3}],  which at LO reconstruct a $\PW$~boson.
Again, the effects are extremely large with $K$~factors reaching ten below the resonance.
The distribution in the invariant mass of the system formed by the
reconstructed bottom quark from the leptonically decaying top quark
and the antimuon [\reffi{fig:inv4}]
has been found to be very sensitive to the top-quark
mass as it possesses an on-shell edge at
$M^2_{\mu^+\bj}<\Mt^2-\MW^2\approx(153\GeV)^2$
\cite{Denner:2012yc,Heinrich:2017bqp}. While the relative corrections
are flat in the on-shell region, they strongly increase above the
on-shell edge.

Finally, \reffis{fig:cos1}--\ref{fig:cos2} show distributions in the
cosine of the production angle of the reconstructed top quarks.
Both distributions are relatively similar up to a reflection of the directions.
Indeed, given that, as opposed to the LHC, the initial state is
asymmetric, the top and antitop quarks have preferred directions
while generally ending up in a back-to-back configuration. 
The NLO QCD corrections are flat and reproduce those of the fiducial
cross section.

\subsection[Differential distributions at $1.5\TeV$]
{Differential distributions at $\mathbf{1.5}\TeV$}
\label{sec:diff1500}

\begin{figure*}
  \centering
  \subfigure[\label{fig:highEn1}]{\includegraphics[width=0.41\textwidth]{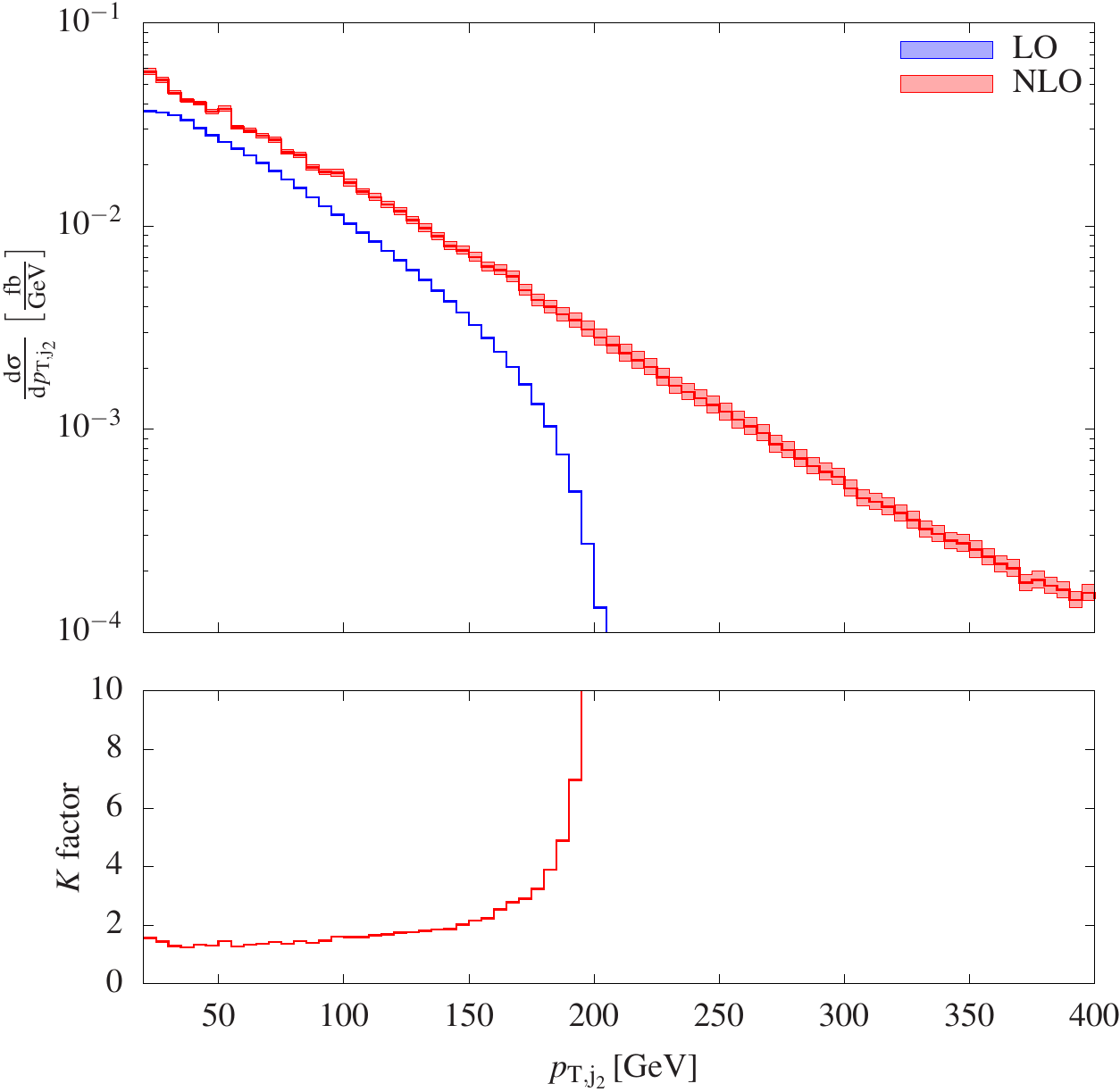}}
  \subfigure[\label{fig:highEn2}]{\includegraphics[width=0.41\textwidth]{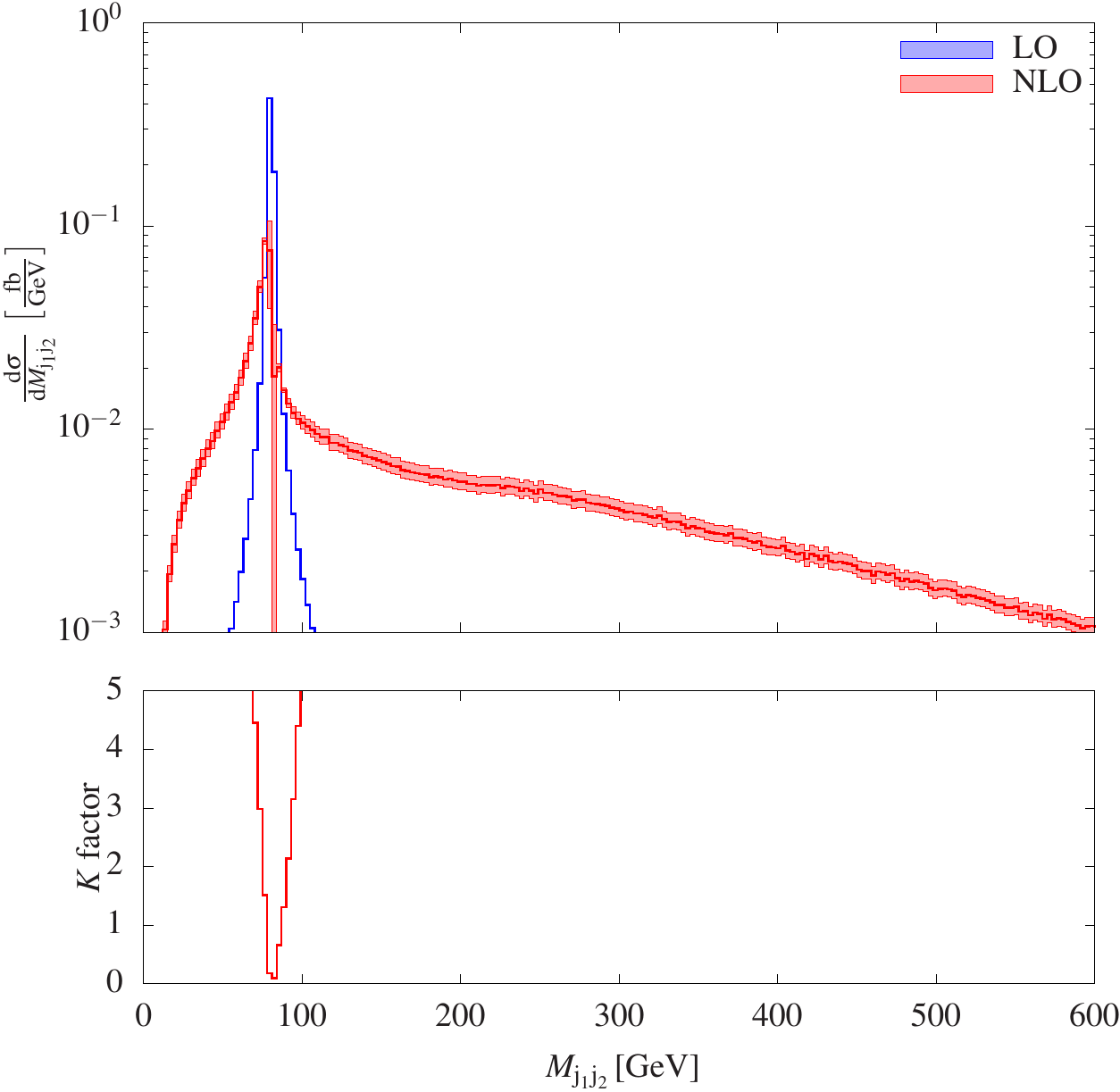}}\\
  \subfigure[\label{fig:highEn3}]{\includegraphics[width=0.42\textwidth]{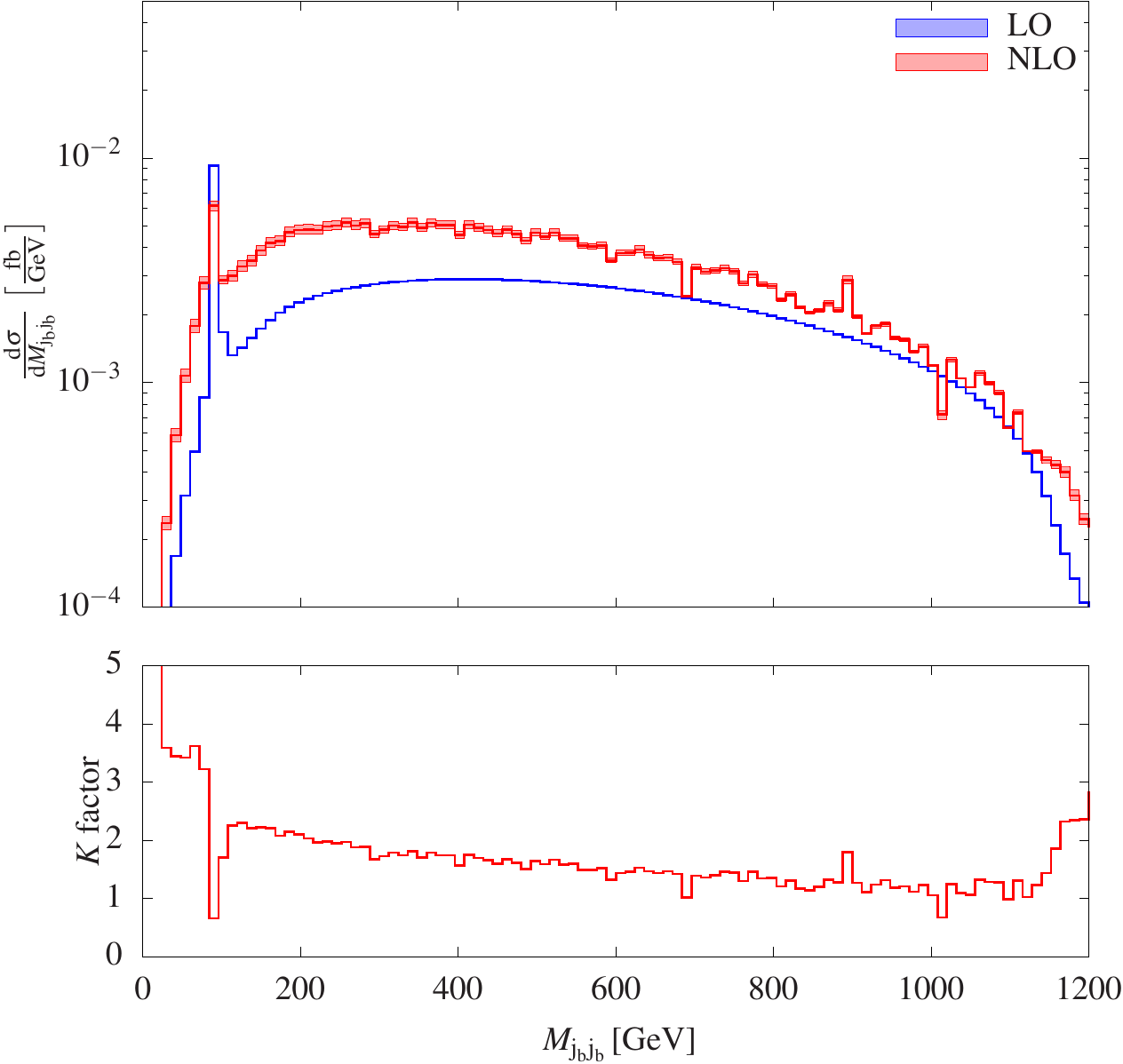}}
  \subfigure[\label{fig:highEn4}]{\includegraphics[width=0.41\textwidth]{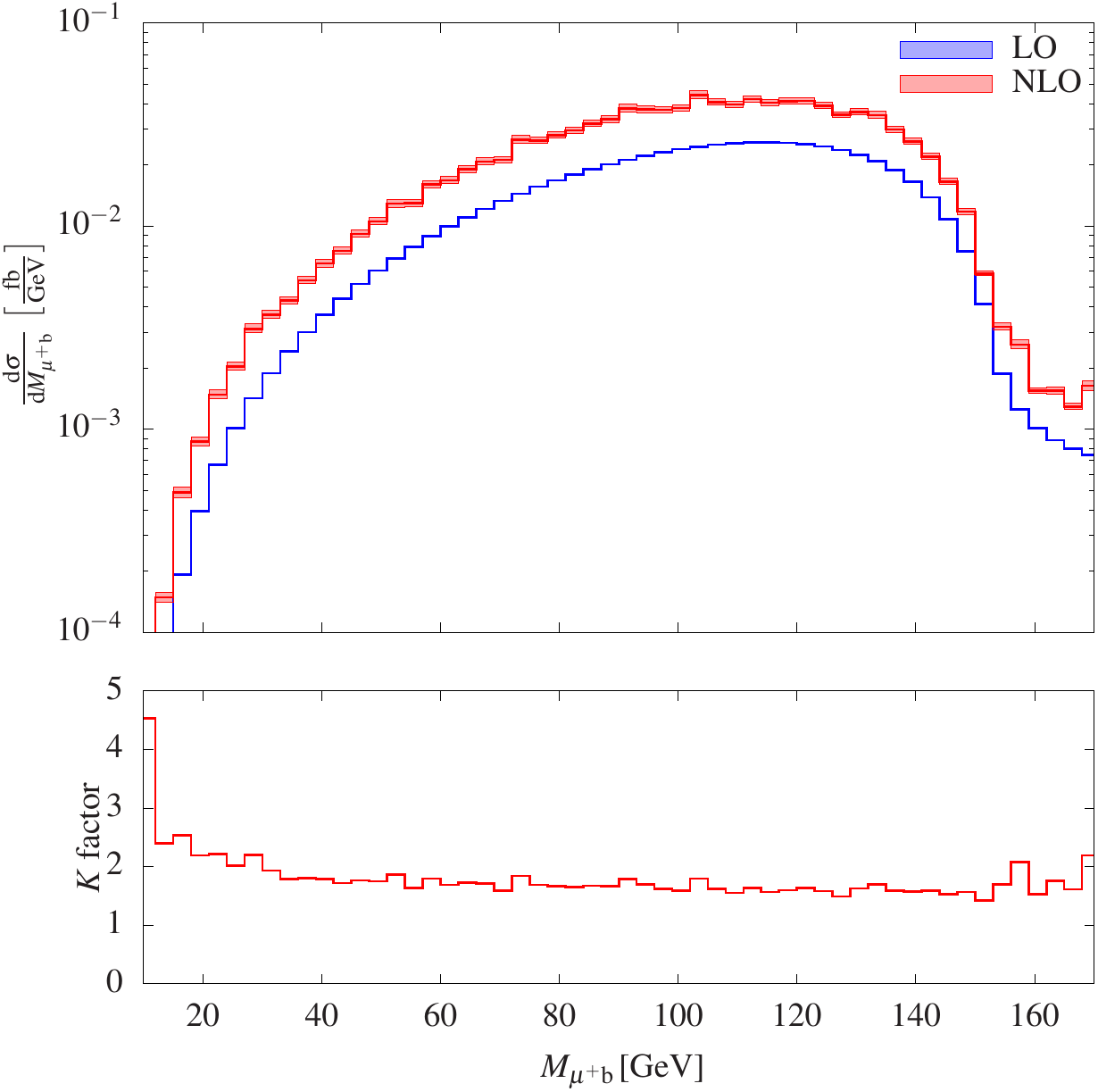}}
  \caption{ 
    Differential distributions for a $1.5\TeV$ CM energy:
    transverse momentum of the second hardest light jet (a),
    invariant masses of the system formed by
    the two hardest light jets (b),
    the two b jets (c), and
    the bottom and the antimuon (d).
    The bottom momentum used in \reffi{fig:highEn4} is taken from Monte Carlo truth.
    Same structure as in \reffi{fig:pT}.}
    \label{fig:highEn}
\end{figure*}

In \reffi{fig:highEn} we show the results obtained for a few selected
observables in $\Pe^+\Pe^-$ collisions at $1.5\TeV$ CM energy.
We stress that the technique used in the $365\GeV$ analysis to reconstruct the top and antitop quarks is not
performing well at $1.5\TeV$, owing to the presence of sizeable
irreducible backgrounds not involving a top--antitop pair 
in the fiducial volume.
Therefore, we do not show top-reconstructed observables in this section as their physical interpretation is unclear.
The higher CM energy shifts the most populated transverse-momentum and invariant-mass
regions to higher values than at $365\GeV$ and enhances irreducible backgrounds that are suppressed for CM energies around the top-pair threshold.

In \reffi{fig:highEn1} we show the differential results in the transverse momentum
of the second-hardest light jet.
At LO and for on-shell $\PW$~bosons, this observable is characterised
by a kinematic cut-off which, assuming small angles between the two
jets, is given by $\pt{\Pj_2,{\rm max}}\sim m_{\Pj\Pj,{\rm max}} / \Delta R_{\Pj\Pj,{\rm min}} \sim\MW/0.4 \sim 200\GeV$~\cite{Denner:2017kzu}.
The NLO QCD corrections, rather flat in the most populated region ($\pt{\Pj_2}\lesssim 150\GeV$),
fill the kinematic regime that is suppressed at LO, driven by real corrections with a third jet from gluon radiation that are tagged as the second-hardest jet.

The light-jet kinematics is strongly affected by QCD corrections at NLO, as can also be observed
in \reffi{fig:highEn2} where the invariant mass of the two hardest light jets is considered.
The typical Breit--Wigner shape coming from the LO $\PW$-boson hadronic decay is distorted by
QCD corrections both below and above the $\PW$-boson pole mass.
The QCD corrections at the peak are negative and qualitatively similar to the ones at $365\GeV$ in \reffi{fig:inv2}.
The radiative tail below $\MW$ is similar to the one observed at $365\GeV$,
while the enhanced $K$~factor found for $M_{\Pj\Pj}\gtrsim 200\GeV$
originates from hard gluon radiation.

The distribution in the invariant mass of the b-jet pair is considered in \reffi{fig:highEn3}.
A clear peak at $\MZ$ highlights the contributions with a resonant
$\PZ$~boson decaying into two b~jets, produced in association with two $\PW$~bosons.
The $\PZ$-boson peak sits on top of contributions coming both from the
$\Pt\bar\Pt$-production process and other sizeable backgrounds as single-top production.
The QCD corrections are large and negative at the $\PZ$-boson peak, similarly
to what can be observed in \reffi{fig:highEn2} for the hadronic $\PW$-boson decay.
Below $\MZ$, the QCD corrections are large and positive, due to the LO suppression
and the presence of unclustered gluon radiation. Above $\MZ$ the corrections
are positive and diminish in size from $100\%$ to a few percent around $1\TeV$.
For an invariant mass close to the maximal possible value for on-shell
production of about $1170\GeV$ [twice the transverse momentum
resulting from Eq.~\refeq{eq:pTjblimit}] the leading order is suppressed
and the $K$~factor increases.

Rather flat relative QCD corrections are found for the distribution in
the invariant mass of the
bottom--antimuon system, shown in \reffi{fig:highEn4}.
As in \reffi{fig:inv4}, the bottom momentum is obtained from Monte Carlo truth.
The edge at $\sqrt{\Mt^2-\MW^2}$ observed in the $365\GeV$ scenario
is present also in the high-energy scenario, but the drop of the LO cross section
around this threshold is less severe at TeV-scale energies
due to the increased irreducible-background contributions which do not embed a resonant
top quark. This leads to a QCD $K$~factor that does not increase for $M_{\mu^+\Pb} > 153\GeV$.

\subsection{Polarised-beam effects}

In the baseline FCC-ee scenarios, the
beams are planned to be unpolarised \cite{FCC:2018evy,Blondel:2019yqr}, while
the CLIC and ILC facilities are envisoned to collide $80\%$-polarised electrons
and possibly $30\%$-polarised positrons \cite{Linssen:2012hp,ILC:2013jhg}.
It has been claimed \cite{Bambade:2019fyw} that polarised beams
at lepton colliders are beneficial to enhance the sensitivity to EW parameters and
possible new-physics effects, increase the signal-to-background ratio
for several signatures, and keep systematics under control.
Assuming a partial polarisation along the beam axis
(often dubbed \emph{longitudinal polarisation})
for both the positron $(P_{\Pe^+})$ and the electron $(P_{\Pe^-})$,
the cross section for a given process reads,
\begin{align}\label{eq:polcombin}
\sigma(P_{\Pe^+}, P_{\Pe^-})\,=\,
\frac14\,
\Bigl[
& \left(1+P_{\Pe^+}\right)(1-P_{\Pe^-})\,\sigma_{\rm RL}\\
+&\left(1-P_{\Pe^+}\right)(1+P_{\Pe^-})\,\sigma_{\rm LR}\nnb\\[0.17cm]
+&\left(1+P_{\Pe^+}\right)(1+P_{\Pe^-})\,\sigma_{\rm RR}\nnb\\
+&\left(1-P_{\Pe^+}\right)(1-P_{\Pe^-})\,\sigma_{\rm LL}\Bigr]\,,\nnb
\end{align}
where $\sigma_{XY}$ is the cross section for a positron with helicity $X$ and an electron with helicity $Y$,
and $\rm L(R)$ stands for left(right)-handed helicity.
Note that in annihilation processes, like the one we consider in this work,
the Standard-Model dynamics only allows for a combined angular momentum equal to $1$. Therefore, the initial-state leptons cannot carry the same helicity, \ie $\sigma_{\rm LL}=\sigma_{\rm RR}=0$.

In \refta{tab:polbeams} we show the integrated cross sections for a number of
\begin{table*}
  \begin{center}
    \begin{tabular}{cc|ccc|ccc}
      &  & \multicolumn{3}{c|}{$\sqrt{s}=365\GeV$} & \multicolumn{3}{c}{$\sqrt{s}=1.5\TeV$}
     \rule[-1ex]{0ex}{2.5ex}\\
       \hline\rule{0ex}{2.7ex}
      $P_{\Pe^+}$ & $P_{\Pe^-}$ & $\sigma_{\rm LO}$ [$\fb$] &
      $\sigma_{\rm NLO \, QCD}$ [$\fb$] & $K$~factor& $\sigma_{\rm
        LO}$ [$\fb$] & $\sigma_{\rm NLO \, QCD}$ [$\fb$] & $K$~factor
      \rule[-1ex]{0ex}{2.5ex}\\
      \hline\rule{0ex}{3ex}
        $0   $ & $    0$ &  $ 16.877(4)  $ & $21.42(1)  ^{+2.23\%}_{ -1.84\%}$ & 1.27 & $2.3235(8)$ & $3.627(9) ^{  +3.77\%}_{ -3.11\%} $ & 1.56 \\[0.1cm]
        $+1   $ & $  -1$ &  $ 48.013(9)  $ & $60.87(4)  ^{+2.22\%}_{ -1.83\%}$ & 1.27 & $7.427(3) $ & $11.09(4) ^{  +3.47\%}_{ -2.86\%} $ & 1.49 \\[0.1cm]
        $-1  $ & $   +1$ &  $ 19.501(2)  $ & $24.86(2)  ^{+2.26\%}_{ -1.86\%}$ & 1.28 & $1.866(1) $ & $3.416(8) ^{  +4.76\%}_{ -3.92\%} $ & 1.83 \\[0.1cm]
        $0   $ & $ -0.8$ &  $ 22.581(5)  $ & $28.63(2)  ^{+2.22\%}_{ -1.83\%}$ & 1.27 & $3.435(2) $ & $5.16(2)  ^{  +3.51\%}_{ -2.89\%} $ & 1.50 \\[0.1cm]
        $0 $   & $+0.8$  &  $ 11.176(3)  $ & $14.23(1)  ^{+ 2.25\%}_{-1.86\%}$ & 1.27 & $1.211(1) $ & $2.091(5) ^{  +4.42\%}_{ -3.64\%} $ & 1.73 \\[0.1cm]
        $-0.3$ & $ +0.8$ &  $ 13.088(3)  $ & $16.67(1)  ^{+2.26\%}_{ -1.86\%}$ & 1.27 & $1.352(1) $ & $2.387(5) ^{  +4.55\%}_{ -3.75\%}$  & 1.77 \\[0.1cm]
        $+0.3 $ & $-0.8$ &  $ 28.770(5)  $ & $36.48(3)  ^{+2.22\%}_{ -1.83\%}$ & 1.27 & $4.410(2) $ & $6.61(2)  ^{  +3.49\%}_{ -2.88\%} $ & 1.50 
    \end{tabular}
  \end{center}
  \caption{\label{tab:polbeams}
    Fiducial cross sections for $\eettsl$ at LO and NLO QCD for
    several beam polarisations and
    for CM energies $365\GeV$ and $1.5\TeV$. The digit in parentheses indicates the Monte
    Carlo statistical error, while the sub- and super-scripts in per cent indicate the scale variation.}
\end{table*}
beam-polarisation choices and for both energy scenarios considered in this work.
Close to the threshold, the configuration with a right-handed positron
and a left-handed electron (RL) gives a fiducial cross section
approximately 2.5 times larger than the opposite helicity
configuration (LR). The QCD corrections relative to the
corresponding LO cross section are almost identical in all
pure or mixed helicity configurations, in agreement with the results
of \citere{ChokoufeNejad:2016qux} for undecayed top quarks. 
At high energy, the ratio between the RL and LR cross sections sizeably
increases at LO, being $\approx 4$ at $1.5\TeV$, while the relative NLO QCD corrections
are larger for the LR state ($+80\%$) than for the RL one ($+50\%$). This
difference between the two pure polarisation states is driven by the real-radiation
contributions which open up new helicity configurations in the final state
that are suppressed at LO.
When irreducible-background effects become relevant, the trivial factorisation
of QCD corrections from the initial-state helicity configuration does not hold anymore \cite{ChokoufeNejad:2016qux}.

\begin{figure*}
  \centering
  \subfigure[\label{fig:pol2_tt}]{\includegraphics[width=0.45\textwidth]{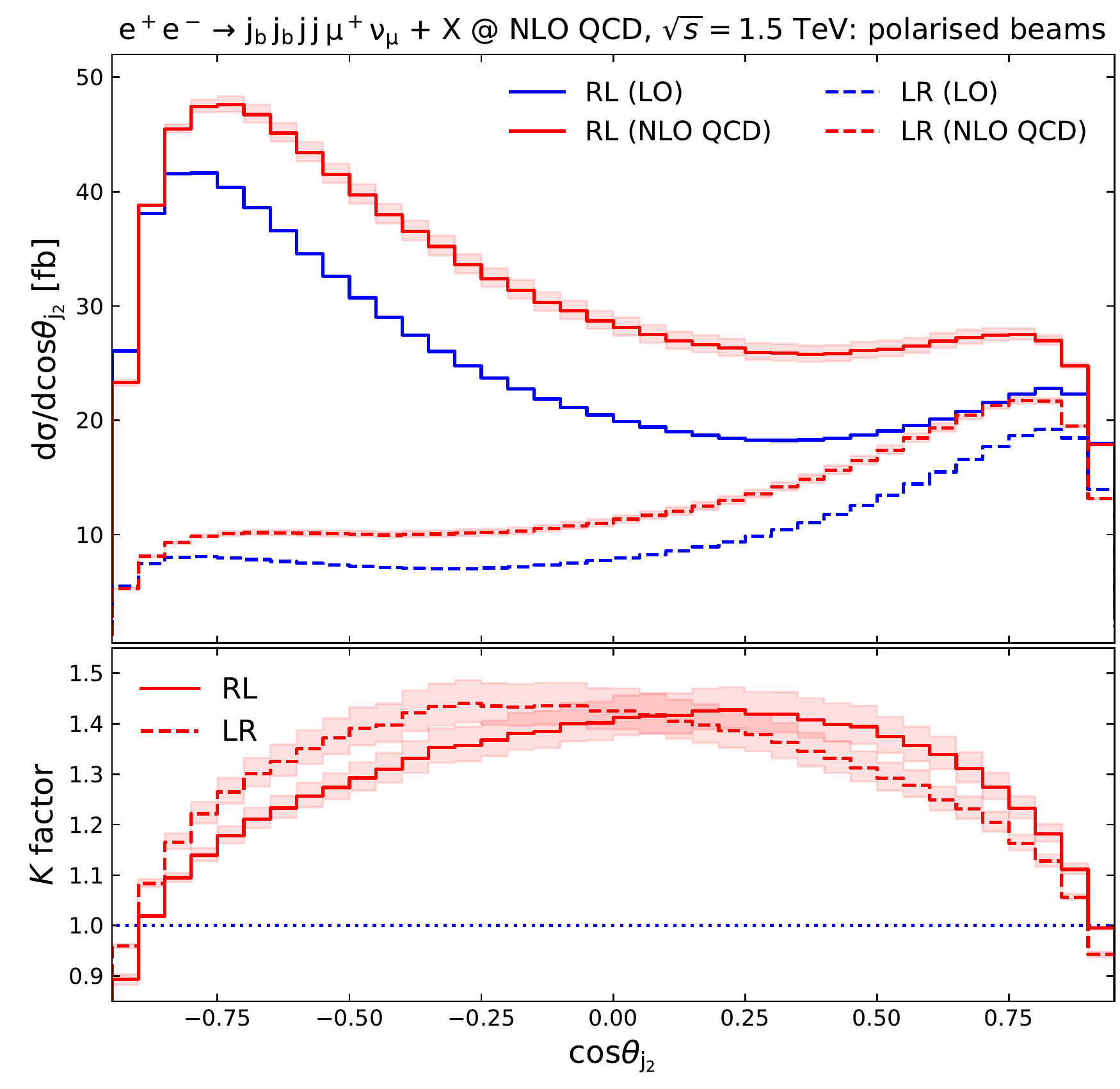}}
  \subfigure[\label{fig:pol2_he}]{\includegraphics[width=0.45\textwidth]{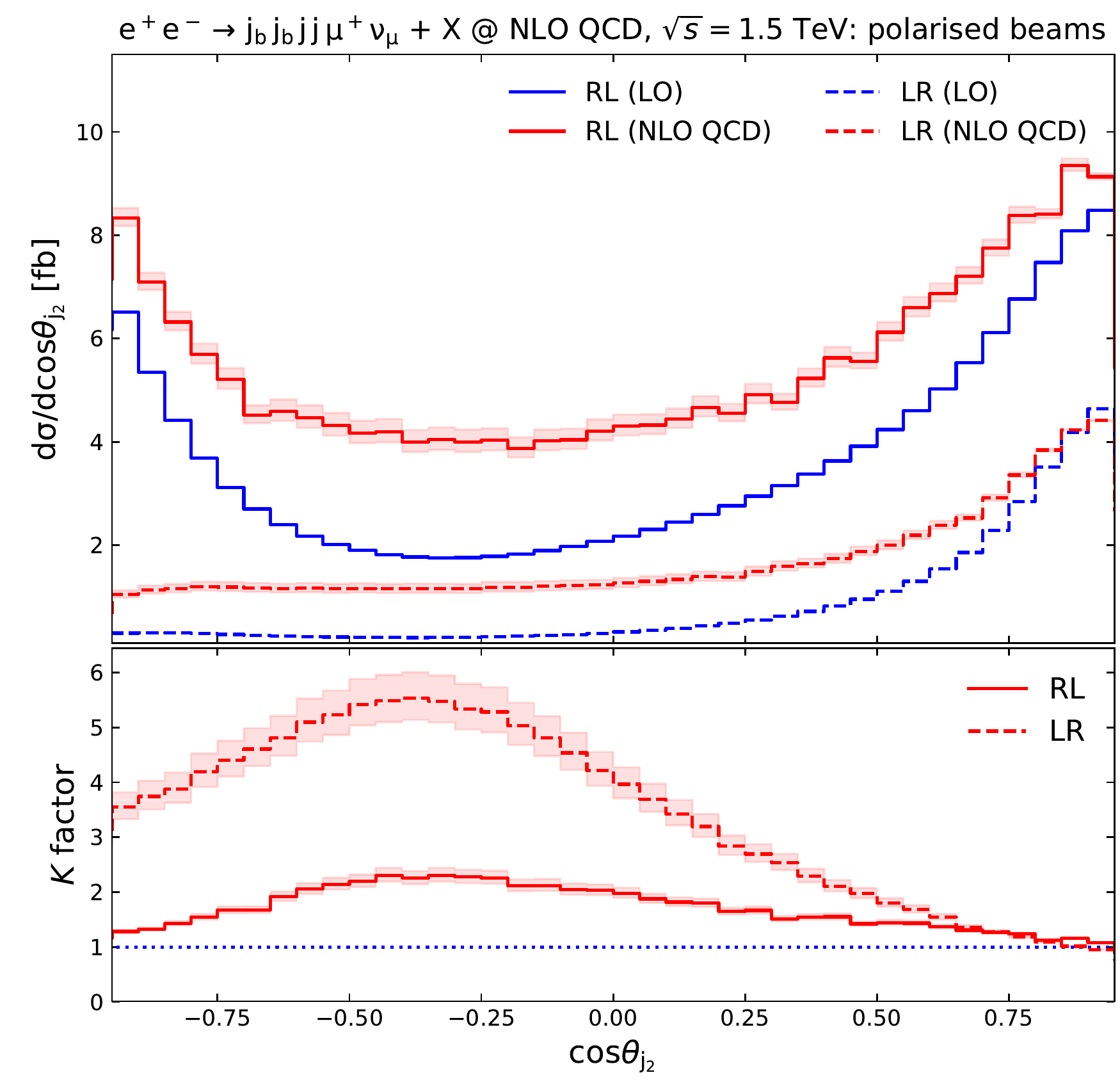}}
  \caption{
    Differential distributions in
    the cosine of the polar angle of the second hardest light jet
    for CM energies $365\GeV$ (left) and $1.5\TeV$ (right).
    The top panels show the absolute differential cross sections at LO (blue) and at NLO QCD (red) for
    the RL (solid) and LR (dashed) helicity combinations of the initial-state leptons.
    The red-shaded band is obtained by means of three-point renormalisation-scale variations and the lower panel displays the $K$~factor.}\label{fig:pol_1}
\end{figure*}

\begin{figure*}
  \centering
  \subfigure[\label{fig:pol4_tt}]{\includegraphics[width=0.45\textwidth]{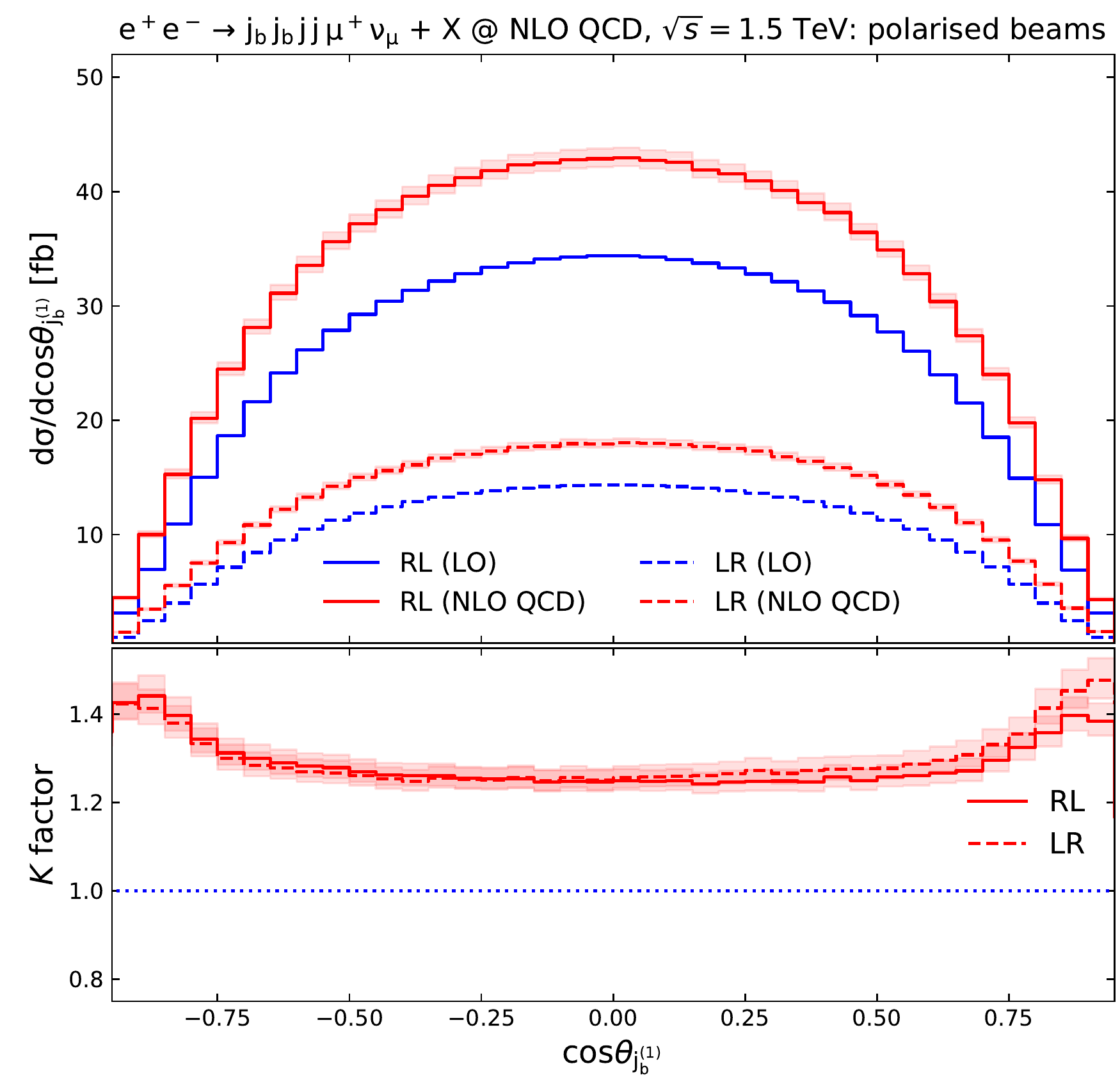}}
  \subfigure[\label{fig:pol4_he}]{\includegraphics[width=0.45\textwidth]{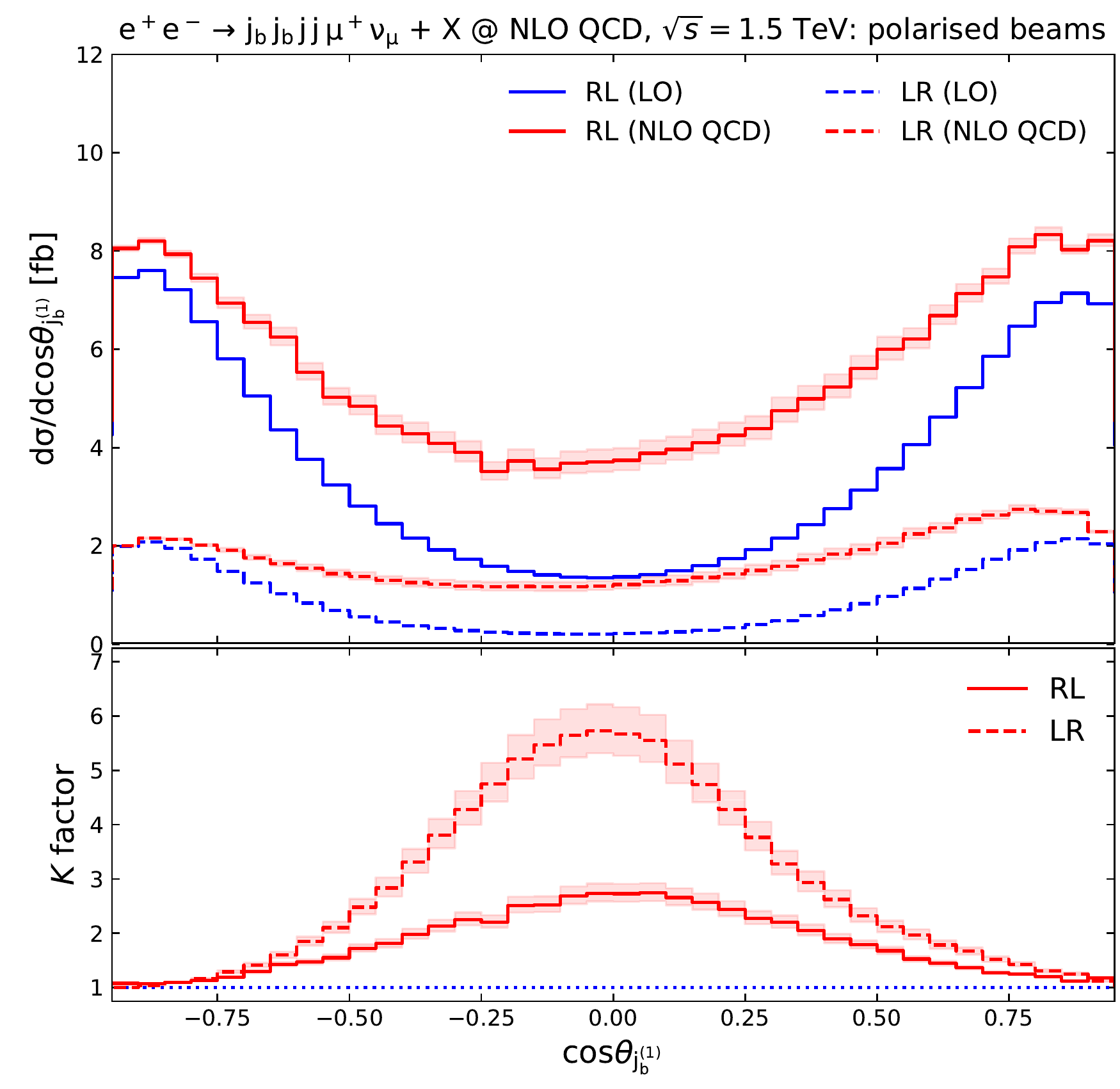}}
  \caption{
    Differential distributions in
    the cosine of the polar angle of the hardest b jet
    for CM energies $365\GeV$ (left) and $1.5\TeV$ (right).
    Same structure as in \reffi{fig:pol_1}.}\label{fig:pol_2}
\end{figure*}

\begin{figure*}
  \centering
  \subfigure[\label{fig:pol3_tt}]{\includegraphics[width=0.45\textwidth]{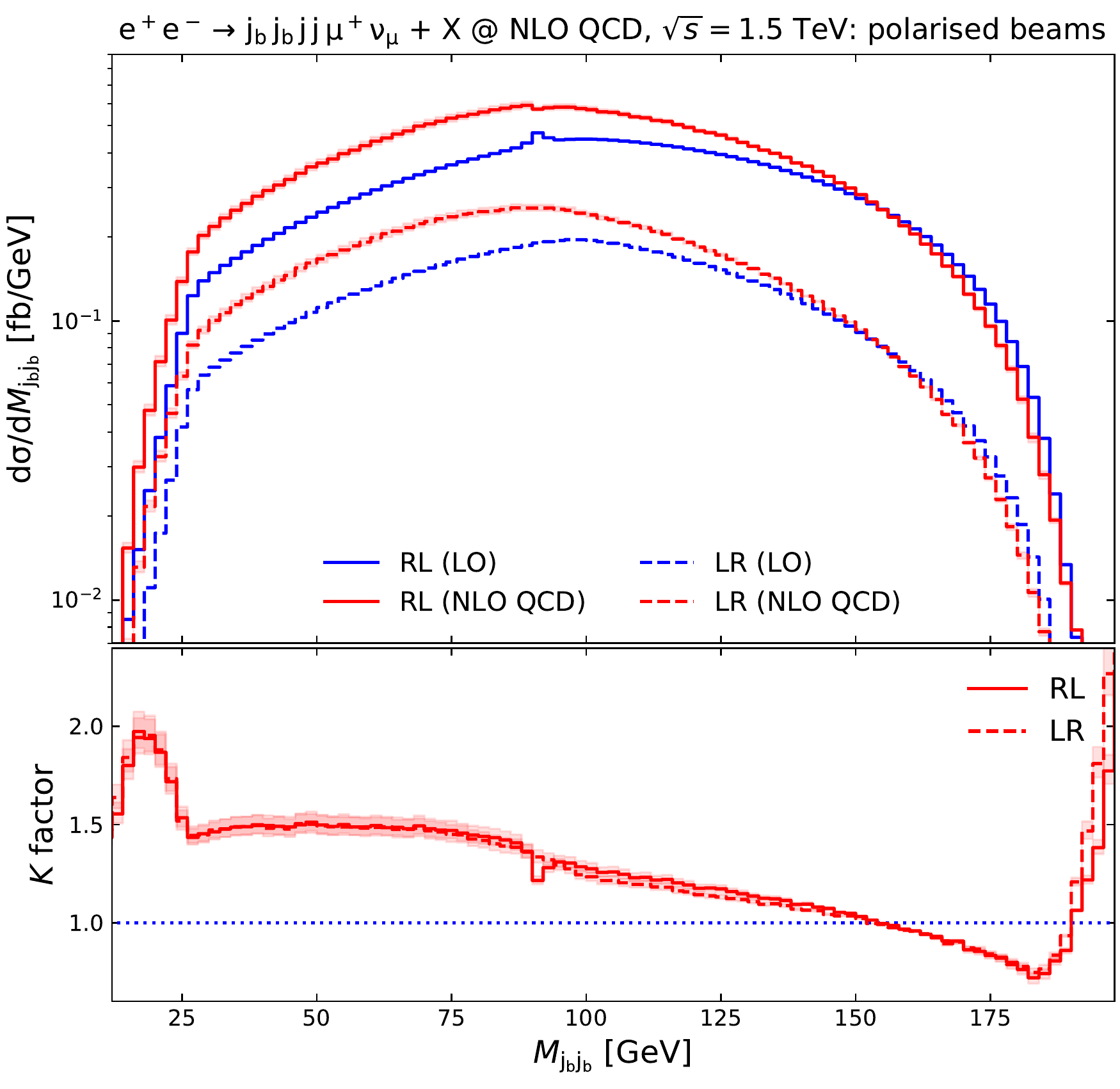}}
  \subfigure[\label{fig:pol3_he}]{\includegraphics[width=0.45\textwidth]{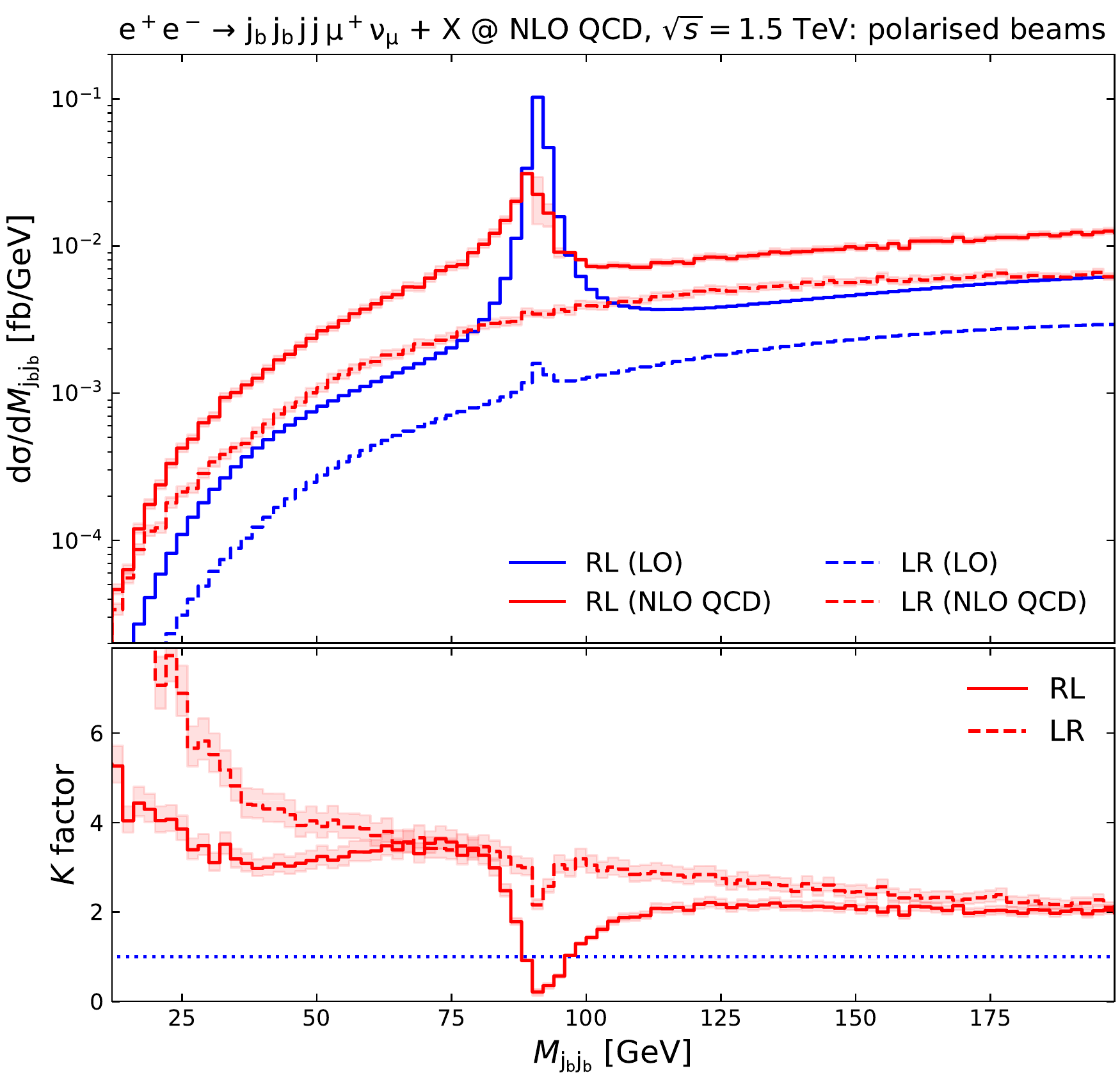}}
  \caption{
    Differential distributions in
    the invariant mass of the b-jet pair
    for CM energies $365\GeV$ (left) and $1.5\TeV$ (right).
    Same structure as in \reffi{fig:pol_1}.}\label{fig:pol_3}
\end{figure*}

The effects of beam polarisation are maximal in differential
distributions, mostly for angular observables.
In \reffis{fig:pol_1}--\ref{fig:pol_3} we show differential results at fixed
initial-state helicities (LR, RL) for both the $365\GeV$ and the $1.5\TeV$ energy scenarios.
Differential results for partially polarised beams, though not considered here,
can be estimated easily performing a bin-by-bin combination of the RL and LR distributions according to
Eq.~\eqref{eq:polcombin}.
At energies well above the $\Pt\bar\Pt$~threshold, selecting the RL helicity state 
is expected to enhance the ratio of single-top and non-resonant
contributions over the top-antitop ones \cite{Fuster:2014hfw}.

In \reffi{fig:pol_1} we consider the distribution in the cosine of the polar angle of the second hardest light jet.
The LR and RL shapes at $365\GeV$ are related by an almost perfect mirroring about $\cos\theta_{\Pj_2}=0$,
up to the different overall normalisation. This holds both at LO and at NLO QCD.
In the RL state, the antitop quark typically goes forward,  as shown
in \reffi{fig:cos2}, and is mostly right handed \cite{Groote:2010zf}.
According to the helicity structure of the tree-level on-shell
top-decay  amplitude, the $\PW^-$~boson from the decay of a mostly right-handed antitop quark
is produced backward with respect to the antitop direction (in the antitop rest frame), therefore giving light jets
that are typically produced with $\cos\theta_{\Pj}<0$. In fact, this
is the case for the second-hardest jet in \reffi{fig:pol2_tt}
but also for the hardest jet, though with different distribution shapes.
The same reasoning applies to the LR initial state, with a flip of sign in
the left--right asymmetry in the decay matrix element, therefore motivating
the almost exact mirror symmetry between the LR and RL states. The QCD
effects are slightly different for the LR and RL states, but in both cases
they are larger where the corresponding LO are more kinematically suppressed.
The situation is significantly different at $1.5\TeV$. The RL state features asymmetric
peaks at the distribution endpoints, the LR one peaks in forward regions, while
being suppressed in backward regions. This results from the fact that
at high energies the antitop quarks have high energies, while still
going mostly in the forward direction for the RL state. As a consequence, their decay
products also often end up in forward direction. 
The QCD corrections mostly fill the
region with negative $\cos\theta_{\Pj_2}$ for both helicity states. Especially for
the LR state, the QCD $K$~factor is between 5 and 6 in the region that is
mostly suppressed at LO.

Shape differences between the LR and RL helicity states are not observed for
all angular observables. In \reffi{fig:pol_2} we show distributions in the cosine of
the polar angle of the hardest b~jet. 
For this variable, the LR and RL distributions are characterised by
very similar LO shapes and QCD $K$~factors at $365\GeV$, with the
most populated region being the central one.
At $1.5\TeV$ the central region is suppressed, while the forward--backward ones
are favoured by both the $\Pt\bar\Pt$ and the single-top
contributions.\footnote{This statement has been verified by
  investigating on-shell top amplitudes obtained from
  \madgraphnlo~\cite{Alwall:2014hca}. Qualitatively it can be
  understood from the fact that both top and antitop quarks are
  preferably produced in the backward and forward directions and thus
  at high energies also their decay products.}
The inclusion of QCD corrections enhances mostly the suppressed
central regions. In spite of a much larger $K$~factor for the LR state,
the NLO shapes are quite similar for the two pure helicity states.
The rationale is that an approximate factorisation of QCD corrections w.r.t.
the beam helicity is fulfilled in the fiducial volume for a beam energy
around the top-mass threshold, while it is broken for TeV-scale CM energies
mostly due to the LO suppression of some helicity configurations and the
opening of new ones thanks to gluon emission.

The difference between different helicity selections can be appreciated not only
in angular distributions, but also in the invariant mass of the
b-jet pair considered in \reffi{fig:pol_3}.
Note that we focus here on masses below $200\GeV$ for both CM energies,
while the unpolarised distribution in \reffi{fig:highEn3} clearly shows that the relevant
range for the $1.5\TeV$ scenario is much larger.
The contributions involving the $\PZ$~decay into two b jets give a peak at the $\PZ$ mass
that is sizeable only in the RL shape, while its presence is hardly
visible for the LR state, since 
 $\Pe^+\Pe^-\rightarrow \PW^+\PW^-\PZ$ contributes basically only for the former
 owing to the  purely left-handed coupling of the $\PW$~boson. While this holds for both energies,
the tri-boson contribution to the cross section is small compared to the dominant $\Pt\bar\Pt$-production process at $365\GeV$, but its contribution is larger at $1.5\TeV$. The difference between the LR and RL
states at $1.5\TeV$ is propagated to the QCD corrections which are sizeable and negative at the $\PZ$~peak
for the RL state, while they are less pronounced for the LR one. On
the contrary, at $365\GeV$, 
the QCD $K$~factors are almost identical for the two helicity states,
apart from a narrow region around the $\PZ$ peak and close to the
kinematic boundary for on-shell top production
$M_{\Pj_\Pb\Pj_\Pb}\lesssim189\GeV$ given by Eq.~\refeq{eq:pTjblimit}
and $M_{\Pj_\Pb\Pj_\Pb}<2p_{\rT,\Pj_\Pb}$.

As shown above, the beam-polarisation effects are relevant for some observables but completely irrelevant for others.
These effects are in general enhanced for higher CM energies, where the interplay
among different subprocesses enhances spin configurations in the final state depending
on the polarisation of the incoming beams.
In particular, the appearance of hard-gluon radiation at NLO can allow for different spin configurations than at LO.

A due comment concerns the flavour of the final-state lepton. Since we have considered $\eettsl$,
with different lepton flavours in the initial and final state no $t$-channel
or boson-fusion topology appears. 
If a positron was considered in the final state, contributions from the
RR and LL initial-state configurations would be non vanishing, and more resonant structures
would be present making the spin structure of the process even more involved.


\section{Conclusion}
\label{sec:con}

Top-quark physics will play a central role in any of the potential future lepton colliders.
It is therefore very important to provide precise theoretical predictions.
So far, only the off-shell production of a top--antitop pair with fully leptonic decays had been computed at NLO QCD.

Nonetheless, the semi-leptonic channel offers many advantages such as the
larger cross section and the possibility to fully reconstruct the
momenta of the top quarks. In this paper, we have provided the first calculation
of the NLO QCD corrections to the full process $\eettsl$.
To that end, we have implemented the FKS subtraction scheme
in the Monte Carlo integrator \mocanlo.  We have successfully validated this
implementation against the already existing Catani--Seymour scheme in
our Monte Carlo program.

At the level of fiducial cross sections, the QCD corrections strongly depend
on the collision energy, ranging from huge positive values at the top-pair threshold
to negative values above threshold but lower than $1\TeV$ and increasingly positive
values beyond $1\TeV$. This dependence results from the Coulomb
singularity in the threshold region and the selection of the decay
jets that suppresses specific kinematic regimes at high energies at LO
but not at NLO because of hard gluon radiation.

The behaviour of QCD corrections becomes even more striking at differential level,
with huge $K$~factors in regions where the LO cross section is
suppressed and the real corrections open up new kinematic topologies.
This holds for both collision energies we have considered ($365\GeV$, $1.5\TeV$).
The radiative corrections become large especially for invariant-mass
and transverse-momentum distributions. 
The most extreme $K$~factors are found at a $1.5\TeV$ collision energy where
the QCD effects are enhanced far from the top-pair threshold.

The structure of LO contributions and QCD radiative corrections can be
understood in more detail when selecting pure helicity states for the
electron and positron beams. The factorisation of higher-order QCD
effects with respect to the spin state of the leptonic initial state
is typically confirmed at collision energies around the top-pair
threshold. It is clearly broken at $1.5\TeV$, owing to enhanced
irreducible background processes and the suppression of certain
helicity configurations at high energy.

\section*{Acknowledgements}

The authors are particularly grateful to Rikkert Frederix and Rene Poncelet for valuable discussions regarding the implementation of the FKS scheme.
The authors thank Maximilian Stahlhofen for useful comments on the manuscript.
A.D.\ and G.P.\ were supported by the German Federal Ministry for Education and Research (BMBF) under contract no.~05H21WWCAA.
M.P.\ acknowledges support by the German Research Foundation (DFG) through the Research Training Group RTG2044 and through grant no INST 39/963-1 FUGG (bwForCluster NEMO) as well as the state of Baden-Württemberg through bwHPC.
The authors acknowledge the hospitality of the Munich Institute for Astro-, Particle and BioPhysics (MIAPbP) where parts of this project were carried out.


\bibliographystyle{elsarticle-num_mod}
\bibliography{fksmocanlo}

\begin{thebibliography}{10}
\expandafter\ifx\csname url\endcsname\relax
  \def\url#1{\texttt{#1}}\fi
\expandafter\ifx\csname urlprefix\endcsname\relax\def\urlprefix{URL }\fi
\expandafter\ifx\csname href\endcsname\relax
  \def\href#1#2{#2} \def\path#1{#1}\fi

\bibitem{ILC:2013jhg}
H.~Baer, et~al. (Eds.), {The International Linear Collider Technical Design
  Report - Volume 2: Physics}, 2013, iLC-REPORT-2013-040.
\newblock \href {http://arxiv.org/abs/1306.6352} {\path{arXiv:1306.6352}}.

\bibitem{Behnke:2013lya}
T.~Behnke, et~al. (Eds.), {The International Linear Collider Technical Design
  Report - Volume 4: Detectors}, 2013, iLC-REPORT-2013-040.
\newblock \href {http://arxiv.org/abs/1306.6329} {\path{arXiv:1306.6329}}.

\bibitem{Bambade:2019fyw}
P.~Bambade, et~al., {The International Linear Collider: A Global Project}
  (2019).
\newblock \href {http://arxiv.org/abs/1903.01629} {\path{arXiv:1903.01629}}.

\bibitem{FCC:2018evy}
A.~Abada, et~al., {FCC-ee: The Lepton Collider}: {Future Circular Collider
  Conceptual Design Report Volume 2}, Eur. Phys. J. ST 228~(2) (2019) 261--623.
\newblock \href {https://doi.org/10.1140/epjst/e2019-900045-4}
  {\path{doi:10.1140/epjst/e2019-900045-4}}.

\bibitem{Linssen:2012hp}
L.~Linssen, A.~Miyamoto, M.~Stanitzki, H.~Weerts (Eds.), {Physics and Detectors
  at CLIC: CLIC Conceptual Design Report}, 2012, {CERN-2012-003}.
\newblock \href {http://arxiv.org/abs/1202.5940} {\path{arXiv:1202.5940}},
  \href {https://doi.org/10.5170/CERN-2012-003}
  {\path{doi:10.5170/CERN-2012-003}}.

\bibitem{Seidel:2013sqa}
K.~Seidel, F.~Simon, M.~Tesar, S.~Poss, {Top quark mass measurements at and
  above threshold at CLIC}, Eur. Phys. J. C 73~(8) (2013) 2530.
\newblock \href {http://arxiv.org/abs/1303.3758} {\path{arXiv:1303.3758}},
  \href {https://doi.org/10.1140/epjc/s10052-013-2530-7}
  {\path{doi:10.1140/epjc/s10052-013-2530-7}}.

\bibitem{Boronat:2016tgd}
M.~Boronat, et~al., {Jet reconstruction at high-energy
  electron\textendash{}positron colliders}, Eur. Phys. J. C 78~(2) (2018) 144.
\newblock \href {http://arxiv.org/abs/1607.05039} {\path{arXiv:1607.05039}},
  \href {https://doi.org/10.1140/epjc/s10052-018-5594-6}
  {\path{doi:10.1140/epjc/s10052-018-5594-6}}.

\bibitem{CLICdp:2018esa}
H.~Abramowicz, et~al., {Top-Quark Physics at the CLIC Electron-Positron Linear
  Collider}, JHEP 11 (2019) 003.
\newblock \href {http://arxiv.org/abs/1807.02441} {\path{arXiv:1807.02441}},
  \href {https://doi.org/10.1007/JHEP11(2019)003}
  {\path{doi:10.1007/JHEP11(2019)003}}.

\bibitem{Dannheim:2019rcr}
D.~Dannheim, K.~Kr\"uger, A.~Levy, A.~N\"urnberg, E.~Sicking (Eds.), {Detector
  Technologies for CLIC}, 2019, {CERN-2019-001}.
\newblock \href {http://arxiv.org/abs/1905.02520} {\path{arXiv:1905.02520}},
  \href {https://doi.org/10.23731/CYRM-2019-001}
  {\path{doi:10.23731/CYRM-2019-001}}.

\bibitem{Hoang:2010gu}
A.~H. Hoang, C.~J. Reisser, P.~Ruiz-Femenia, {Phase Space Matching and Finite
  Lifetime Effects for Top-Pair Production Close to Threshold}, Phys. Rev. D 82
  (2010) 014005.
\newblock \href {http://arxiv.org/abs/1002.3223} {\path{arXiv:1002.3223}},
  \href {https://doi.org/10.1103/PhysRevD.82.014005}
  {\path{doi:10.1103/PhysRevD.82.014005}}.

\bibitem{Hoang:2013uda}
A.~H. Hoang, M.~Stahlhofen, {The Top-Antitop Threshold at the ILC: NNLL QCD
  Uncertainties}, JHEP 05 (2014) 121.
\newblock \href {http://arxiv.org/abs/1309.6323} {\path{arXiv:1309.6323}},
  \href {https://doi.org/10.1007/JHEP05(2014)121}
  {\path{doi:10.1007/JHEP05(2014)121}}.

\bibitem{Beneke:2015kwa}
M.~Beneke, et~al., {Next-to-Next-to-Next-to-Leading Order QCD Prediction for
  the Top Antitop $S$-Wave Pair Production Cross Section Near Threshold in
  $e^+e^-$ Annihilation}, Phys. Rev. Lett. 115~(19) (2015) 192001.
\newblock \href {http://arxiv.org/abs/1506.06864} {\path{arXiv:1506.06864}},
  \href {https://doi.org/10.1103/PhysRevLett.115.192001}
  {\path{doi:10.1103/PhysRevLett.115.192001}}.

\bibitem{Beneke:2017rdn}
M.~Beneke, A.~Maier, T.~Rauh, P.~Ruiz-Femenia, {Non-resonant and electroweak
  NNLO correction to the $e^+ e^-$ top anti-top threshold}, JHEP 02 (2018) 125.
\newblock \href {http://arxiv.org/abs/1711.10429} {\path{arXiv:1711.10429}},
  \href {https://doi.org/10.1007/JHEP02(2018)125}
  {\path{doi:10.1007/JHEP02(2018)125}}.

\bibitem{Bach:2017ggt}
F.~Bach, et~al., {Fully-differential Top-Pair Production at a Lepton Collider:
  From Threshold to Continuum}, JHEP 03 (2018) 184.
\newblock \href {http://arxiv.org/abs/1712.02220} {\path{arXiv:1712.02220}},
  \href {https://doi.org/10.1007/JHEP03(2018)184}
  {\path{doi:10.1007/JHEP03(2018)184}}.

\bibitem{Hoang:2008qy}
A.~H. Hoang, V.~Mateu, S.~Mohammad~Zebarjad, {Heavy Quark Vacuum Polarization
  Function at ${\cal O}(\alpha^2_s)$ and ${\cal O}(\alpha^3_s)$}, Nucl. Phys. B
  813 (2009) 349--369.
\newblock \href {http://arxiv.org/abs/0807.4173} {\path{arXiv:0807.4173}},
  \href {https://doi.org/10.1016/j.nuclphysb.2008.12.005}
  {\path{doi:10.1016/j.nuclphysb.2008.12.005}}.

\bibitem{Kiyo:2009gb}
Y.~Kiyo, A.~Maier, P.~Maierh{\"o}fer, P.~Marquard, {Reconstruction of heavy
  quark current correlators at ${\cal O}(\alpha_s^3)$}, Nucl. Phys. B 823
  (2009) 269--287.
\newblock \href {http://arxiv.org/abs/0907.2120} {\path{arXiv:0907.2120}},
  \href {https://doi.org/10.1016/j.nuclphysb.2009.08.010}
  {\path{doi:10.1016/j.nuclphysb.2009.08.010}}.

\bibitem{Gao:2014nva}
J.~Gao, H.~X. Zhu, {Electroweak prodution of top-quark pairs in e$^+$e$^-$
  annihilation at NNLO in QCD: the vector contributions}, Phys. Rev. D 90~(11)
  (2014) 114022.
\newblock \href {http://arxiv.org/abs/1408.5150} {\path{arXiv:1408.5150}},
  \href {https://doi.org/10.1103/PhysRevD.90.114022}
  {\path{doi:10.1103/PhysRevD.90.114022}}.

\bibitem{Gao:2014eea}
J.~Gao, H.~X. Zhu, {Top Quark Forward-Backward Asymmetry in $e^+e^-$
  Annihilation at Next-to-Next-to-Leading Order in QCD}, Phys. Rev. Lett.
  113~(26) (2014) 262001.
\newblock \href {http://arxiv.org/abs/1410.3165} {\path{arXiv:1410.3165}},
  \href {https://doi.org/10.1103/PhysRevLett.113.262001}
  {\path{doi:10.1103/PhysRevLett.113.262001}}.

\bibitem{Chen:2016zbz}
L.~Chen, O.~Dekkers, D.~Heisler, W.~Bernreuther, Z.-G. Si, {Top-quark pair
  production at next-to-next-to-leading order QCD in electron positron
  collisions}, JHEP 12 (2016) 098.
\newblock \href {http://arxiv.org/abs/1610.07897} {\path{arXiv:1610.07897}},
  \href {https://doi.org/10.1007/JHEP12(2016)098}
  {\path{doi:10.1007/JHEP12(2016)098}}.

\bibitem{Bernreuther:2023ulo}
W.~Bernreuther, L.~Chen, P.-C. Lu, Z.-G. Si, {Top and bottom quark
  forward-backward asymmetries at next-to-next-to-leading order QCD in
  (un)polarized electron positron collisions} (2023).
\newblock \href {http://arxiv.org/abs/2301.12632} {\path{arXiv:2301.12632}}.

\bibitem{Guo:2008clc}
L.~Guo, W.-G. Ma, R.-Y. Zhang, S.-M. Wang, {One-loop QCD corrections to the
  $e^+ e^-\to W^+ W^- b \bar{b}$ process at the ILC}, Phys. Lett. B 662 (2008)
  150--157.
\newblock \href {http://arxiv.org/abs/0802.4124} {\path{arXiv:0802.4124}},
  \href {https://doi.org/10.1016/j.physletb.2008.02.058}
  {\path{doi:10.1016/j.physletb.2008.02.058}}.

\bibitem{Liebler:2015ipp}
S.~Liebler, G.~Moortgat-Pick, A.~S. Papanastasiou, {Probing the top-quark width
  through ratios of resonance contributions of $e^+e^-\rightarrow
  W^+W^-b\bar{b}$}, JHEP 03 (2016) 099.
\newblock \href {http://arxiv.org/abs/1511.02350} {\path{arXiv:1511.02350}},
  \href {https://doi.org/10.1007/JHEP03(2016)099}
  {\path{doi:10.1007/JHEP03(2016)099}}.

\bibitem{ChokoufeNejad:2016qux}
B.~Chokouf\'e~Nejad, et~al., {NLO QCD predictions for off-shell $ t\overline{t}
  $ and $ t\overline{t}H $ production and decay at a linear collider}, JHEP 12
  (2016) 075.
\newblock \href {http://arxiv.org/abs/1609.03390} {\path{arXiv:1609.03390}},
  \href {https://doi.org/10.1007/JHEP12(2016)075}
  {\path{doi:10.1007/JHEP12(2016)075}}.

\bibitem{Fujimoto:1987hu}
J.~Fujimoto, Y.~Shimizu, {Radiative Corrections to $e^+ e^- \to \bar{t}t$ in
  Electroweak Theory}, Mod. Phys. Lett. A 3 (1988) 581.
\newblock \href {https://doi.org/10.1142/S0217732388000696}
  {\path{doi:10.1142/S0217732388000696}}.

\bibitem{Beenakker:1991ca}
W.~Beenakker, S.~C. van~der Marck, W.~Hollik, {$e^+ e^-$ annihilation into
  heavy fermion pairs at high-energy colliders}, Nucl. Phys. B 365 (1991)
  24--78.
\newblock \href {https://doi.org/10.1016/0550-3213(91)90606-X}
  {\path{doi:10.1016/0550-3213(91)90606-X}}.

\bibitem{Fleischer:2003kk}
J.~Fleischer, A.~Leike, T.~Riemann, A.~Werthenbach, {Electroweak one loop
  corrections for $e^+ e^-$ annihilation into $t\bar{t}$ including hard
  bremsstrahlung}, Eur. Phys. J. C 31 (2003) 37--56.
\newblock \href {http://arxiv.org/abs/hep-ph/0302259}
  {\path{arXiv:hep-ph/0302259}}, \href
  {https://doi.org/10.1140/epjc/s2003-01263-8}
  {\path{doi:10.1140/epjc/s2003-01263-8}}.

\bibitem{NhiMUQuach:2017lrx}
N.~M.~U. Quach, Y.~Kurihara, {ISR effects on loop corrections of a top
  pair-production at the ILC}, J. Phys. Conf. Ser. 920~(1) (2017) 012012.
\newblock \href {http://arxiv.org/abs/1706.07042} {\path{arXiv:1706.07042}},
  \href {https://doi.org/10.1088/1742-6596/920/1/012012}
  {\path{doi:10.1088/1742-6596/920/1/012012}}.

\bibitem{Bertone:2022ktl}
V.~Bertone, et~al., {Improving methods and predictions at high-energy
  e$^+$e$^-$ colliders within collinear factorisation}, JHEP 10 (2022) 089.
\newblock \href {http://arxiv.org/abs/2207.03265} {\path{arXiv:2207.03265}},
  \href {https://doi.org/10.1007/JHEP10(2022)089}
  {\path{doi:10.1007/JHEP10(2022)089}}.

\bibitem{Amjad:2013tlv}
M.~S. Amjad, et~al., {A precise determination of top quark electro-weak
  couplings at the ILC operating at $\sqrt{s}=500$ GeV} (2013).
\newblock \href {http://arxiv.org/abs/1307.8102} {\path{arXiv:1307.8102}}.

\bibitem{Fuster:2014hfw}
J.~Fuster, et~al., {Study of single top production at high energy electron
  positron colliders}, Eur. Phys. J. C 75 (2015) 223.
\newblock \href {http://arxiv.org/abs/1411.2355} {\path{arXiv:1411.2355}},
  \href {https://doi.org/10.1140/epjc/s10052-015-3453-2}
  {\path{doi:10.1140/epjc/s10052-015-3453-2}}.

\bibitem{Amjad:2015mma}
M.~S. Amjad, et~al., {A precise characterisation of the top quark electro-weak
  vertices at the ILC}, Eur. Phys. J. C 75~(10) (2015) 512.
\newblock \href {http://arxiv.org/abs/1505.06020} {\path{arXiv:1505.06020}},
  \href {https://doi.org/10.1140/epjc/s10052-015-3746-5}
  {\path{doi:10.1140/epjc/s10052-015-3746-5}}.

\bibitem{Bernreuther:2017cyi}
W.~Bernreuther, et~al., {CP-violating top quark couplings at future linear
  $e^+e^-$ colliders}, Eur. Phys. J. C 78~(2) (2018) 155.
\newblock \href {http://arxiv.org/abs/1710.06737} {\path{arXiv:1710.06737}},
  \href {https://doi.org/10.1140/epjc/s10052-018-5625-3}
  {\path{doi:10.1140/epjc/s10052-018-5625-3}}.

\bibitem{Denner:2017kzu}
A.~Denner, M.~Pellen, {Off-shell production of top-antitop pairs in the
  lepton+jets channel at NLO QCD}, JHEP 02 (2018) 013.
\newblock \href {http://arxiv.org/abs/1711.10359} {\path{arXiv:1711.10359}},
  \href {https://doi.org/10.1007/JHEP02(2018)013}
  {\path{doi:10.1007/JHEP02(2018)013}}.

\bibitem{Frixione:1995ms}
S.~Frixione, Z.~Kunszt, A.~Signer, {Three jet cross-sections to next-to-leading
  order}, Nucl. Phys. B 467 (1996) 399--442.
\newblock \href {http://arxiv.org/abs/hep-ph/9512328}
  {\path{arXiv:hep-ph/9512328}}, \href
  {https://doi.org/10.1016/0550-3213(96)00110-1}
  {\path{doi:10.1016/0550-3213(96)00110-1}}.

\bibitem{ParticleDataGroup:2020ssz}
P.~A. Zyla, et~al., {Review of Particle Physics}, PTEP 2020~(8) (2020) 083C01.
\newblock \href {https://doi.org/10.1093/ptep/ptaa104}
  {\path{doi:10.1093/ptep/ptaa104}}.

\bibitem{Bardin:1988xt}
D.~Bardin, A.~Leike, T.~Riemann, M.~Sachwitz, {Energy-dependent width effects
  in $e^+ e^-$ annihilation near the Z-boson pole}, Phys. Lett. B 206 (1988)
  539--542.
\newblock \href {https://doi.org/10.1016/0370-2693(88)91627-9}
  {\path{doi:10.1016/0370-2693(88)91627-9}}.

\bibitem{Heinemeyer:2013tqa}
S.~Heinemeyer, C.~Mariotti, G.~Passarino, R.~Tanaka (Eds.), {Handbook of LHC
  Higgs Cross Sections: 3. Higgs Properties}, CERN, Geneva, 2013,
  {CERN-2013-004}.
\newblock \href {http://arxiv.org/abs/1307.1347} {\path{arXiv:1307.1347}},
  \href {https://doi.org/10.5170/CERN-2013-004}
  {\path{doi:10.5170/CERN-2013-004}}.

\bibitem{Basso:2015gca}
L.~Basso, S.~Dittmaier, A.~Huss, L.~Oggero, {Techniques for the treatment of IR
  divergences in decay processes at NLO and application to the top-quark
  decay}, Eur. Phys. J. C76 (2016) 56.
\newblock \href {http://arxiv.org/abs/1507.04676} {\path{arXiv:1507.04676}},
  \href {https://doi.org/10.1140/epjc/s10052-016-3878-2}
  {\path{doi:10.1140/epjc/s10052-016-3878-2}}.

\bibitem{Jezabek:1988iv}
M.~Je\.{z}abek, J.~H. K{\"u}hn, {QCD Corrections to Semileptonic Decays of
  Heavy Quarks}, Nucl. Phys. B314 (1989) 1--6.
\newblock \href {https://doi.org/10.1016/0550-3213(89)90108-9}
  {\path{doi:10.1016/0550-3213(89)90108-9}}.

\bibitem{Denner:1999gp}
A.~Denner, S.~Dittmaier, M.~Roth, D.~Wackeroth, {Predictions for all processes
  $e^+ e^-\to 4\,$fermions${} + \gamma$}, Nucl. Phys. B 560 (1999) 33--65.
\newblock \href {http://arxiv.org/abs/hep-ph/9904472}
  {\path{arXiv:hep-ph/9904472}}, \href
  {https://doi.org/10.1016/S0550-3213(99)00437-X}
  {\path{doi:10.1016/S0550-3213(99)00437-X}}.

\bibitem{Denner:2005fg}
A.~Denner, S.~Dittmaier, M.~Roth, L.~Wieders, {Electroweak corrections to
  charged-current $e^+ e^-\to 4\,$fermion processes: Technical details and
  further results}, Nucl. Phys. B 724 (2005) 247--294, [Erratum: Nucl.\ Phys.\
  B 854 (2012) 504].
\newblock \href {http://arxiv.org/abs/hep-ph/0505042}
  {\path{arXiv:hep-ph/0505042}}, \href
  {https://doi.org/10.1016/j.nuclphysb.2011.09.001}
  {\path{doi:10.1016/j.nuclphysb.2011.09.001}}.

\bibitem{Denner:2006ic}
A.~Denner, S.~Dittmaier, {The complex-mass scheme for perturbative calculations
  with unstable particles}, Nucl. Phys. Proc. Suppl. 160 (2006) 22--26.
\newblock \href {http://arxiv.org/abs/hep-ph/0605312}
  {\path{arXiv:hep-ph/0605312}}, \href
  {https://doi.org/10.1016/j.nuclphysbps.2006.09.025}
  {\path{doi:10.1016/j.nuclphysbps.2006.09.025}}.

\bibitem{Denner:2019vbn}
A.~Denner, S.~Dittmaier, {Electroweak Radiative Corrections for Collider
  Physics}, Phys. Rept. 864 (2020) 1--163.
\newblock \href {http://arxiv.org/abs/1912.06823} {\path{arXiv:1912.06823}},
  \href {https://doi.org/10.1016/j.physrep.2020.04.001}
  {\path{doi:10.1016/j.physrep.2020.04.001}}.

\bibitem{Denner:2000bj}
A.~Denner, S.~Dittmaier, M.~Roth, D.~Wackeroth, {Electroweak radiative
  corrections to ${e}^+ {e}^- \to {W W} \to$ 4 fermions in double-pole
  approximation: The RACOONWW approach}, Nucl. Phys. B587 (2000) 67--117.
\newblock \href {http://arxiv.org/abs/hep-ph/0006307}
  {\path{arXiv:hep-ph/0006307}}, \href
  {https://doi.org/10.1016/S0550-3213(00)00511-3}
  {\path{doi:10.1016/S0550-3213(00)00511-3}}.

\bibitem{Actis:2016mpe}
S.~Actis, et~al., {RECOLA: REcursive Computation of One-Loop Amplitudes},
  Comput. Phys. Commun. 214 (2017) 140--173.
\newblock \href {http://arxiv.org/abs/1605.01090} {\path{arXiv:1605.01090}},
  \href {https://doi.org/10.1016/j.cpc.2017.01.004}
  {\path{doi:10.1016/j.cpc.2017.01.004}}.

\bibitem{CLIC:2016zwp}
M.~J. Boland, et~al., {Updated baseline for a staged Compact Linear Collider},
  2016, {CERN-2016-004}.
\newblock \href {http://arxiv.org/abs/1608.07537} {\path{arXiv:1608.07537}},
  \href {https://doi.org/10.5170/CERN-2016-004}
  {\path{doi:10.5170/CERN-2016-004}}.

\bibitem{Catani:1993hr}
S.~Catani, Y.~L. Dokshitzer, M.~H. Seymour, B.~R. Webber, {Longitudinally
  invariant $K_t$ clustering algorithms for hadron hadron collisions}, Nucl.
  Phys. B 406 (1993) 187--224.
\newblock \href {https://doi.org/10.1016/0550-3213(93)90166-M}
  {\path{doi:10.1016/0550-3213(93)90166-M}}.

\bibitem{Denner:2020orv}
A.~Denner, J.-N. Lang, M.~Pellen, {Full NLO QCD corrections to off-shell ttbb
  production}, Phys. Rev. D 104~(5) (2021) 056018.
\newblock \href {http://arxiv.org/abs/2008.00918} {\path{arXiv:2008.00918}},
  \href {https://doi.org/10.1103/PhysRevD.104.056018}
  {\path{doi:10.1103/PhysRevD.104.056018}}.

\bibitem{Denner:2015yca}
A.~Denner, R.~Feger, {NLO QCD corrections to off-shell top-antitop production
  with leptonic decays in association with a Higgs boson at the LHC}, JHEP 11
  (2015) 209.
\newblock \href {http://arxiv.org/abs/1506.07448} {\path{arXiv:1506.07448}},
  \href {https://doi.org/10.1007/JHEP11(2015)209}
  {\path{doi:10.1007/JHEP11(2015)209}}.

\bibitem{Denner:2016jyo}
A.~Denner, M.~Pellen, {NLO electroweak corrections to off-shell top-antitop
  production with leptonic decays at the LHC}, JHEP 08 (2016) 155.
\newblock \href {http://arxiv.org/abs/1607.05571} {\path{arXiv:1607.05571}},
  \href {https://doi.org/10.1007/JHEP08(2016)155}
  {\path{doi:10.1007/JHEP08(2016)155}}.

\bibitem{Denner:2016wet}
A.~Denner, J.-N. Lang, M.~Pellen, S.~Uccirati, {Higgs production in association
  with off-shell top-antitop pairs at NLO EW and QCD at the LHC}, JHEP 02
  (2017) 053.
\newblock \href {http://arxiv.org/abs/1612.07138} {\path{arXiv:1612.07138}},
  \href {https://doi.org/10.1007/JHEP02(2017)053}
  {\path{doi:10.1007/JHEP02(2017)053}}.

\bibitem{Denner:2020hgg}
A.~Denner, G.~Pelliccioli, {NLO QCD corrections to off-shell
  $\text{t}\bar{\text{t}}\text{W}^+$ production at the LHC}, JHEP 11 (2020)
  069.
\newblock \href {http://arxiv.org/abs/2007.12089} {\path{arXiv:2007.12089}},
  \href {https://doi.org/10.1007/JHEP11(2020)069}
  {\path{doi:10.1007/JHEP11(2020)069}}.

\bibitem{Denner:2021hqi}
A.~Denner, G.~Pelliccioli, {Combined NLO EW and QCD corrections to off-shell
  $\text {t} \overline{\text {t}}\text {W} $ production at the LHC}, Eur. Phys.
  J. C 81~(4) (2021) 354.
\newblock \href {http://arxiv.org/abs/2102.03246} {\path{arXiv:2102.03246}},
  \href {https://doi.org/10.1140/epjc/s10052-021-09143-3}
  {\path{doi:10.1140/epjc/s10052-021-09143-3}}.

\bibitem{Denner:2022fhu}
A.~Denner, G.~Pelliccioli, C.~Schwan, {NLO QCD and EW corrections to off-shell
  tZj production at the LHC}, JHEP 10 (2022) 125.
\newblock \href {http://arxiv.org/abs/2207.11264} {\path{arXiv:2207.11264}},
  \href {https://doi.org/10.1007/JHEP10(2022)125}
  {\path{doi:10.1007/JHEP10(2022)125}}.

\bibitem{Berends:1994pv}
F.~A. Berends, R.~Pittau, R.~Kleiss, {All electroweak four fermion processes in
  electron--positron collisions}, Nucl. Phys. B 424 (1994) 308--342.
\newblock \href {http://arxiv.org/abs/hep-ph/9404313}
  {\path{arXiv:hep-ph/9404313}}, \href
  {https://doi.org/10.1016/0550-3213(94)90297-6}
  {\path{doi:10.1016/0550-3213(94)90297-6}}.

\bibitem{Dittmaier:2002ap}
S.~Dittmaier, M.~Roth, {LUSIFER: A LUcid approach to six FERmion production},
  Nucl. Phys. B 642 (2002) 307--343.
\newblock \href {http://arxiv.org/abs/hep-ph/0206070}
  {\path{arXiv:hep-ph/0206070}}, \href
  {https://doi.org/10.1016/S0550-3213(02)00640-5}
  {\path{doi:10.1016/S0550-3213(02)00640-5}}.

\bibitem{Actis:2012qn}
S.~Actis, A.~Denner, L.~Hofer, A.~Scharf, S.~Uccirati, {Recursive generation of
  one-loop amplitudes in the Standard Model}, JHEP 04 (2013) 037.
\newblock \href {http://arxiv.org/abs/1211.6316} {\path{arXiv:1211.6316}},
  \href {https://doi.org/10.1007/JHEP04(2013)037}
  {\path{doi:10.1007/JHEP04(2013)037}}.

\bibitem{Denner:2016kdg}
A.~Denner, S.~Dittmaier, L.~Hofer, {COLLIER: a fortran-based Complex One-Loop
  LIbrary in Extended Regularizations}, Comput. Phys. Commun. 212 (2017)
  220--238.
\newblock \href {http://arxiv.org/abs/1604.06792} {\path{arXiv:1604.06792}},
  \href {https://doi.org/10.1016/j.cpc.2016.10.013}
  {\path{doi:10.1016/j.cpc.2016.10.013}}.

\bibitem{Catani:1996vz}
S.~Catani, M.~Seymour, {A general algorithm for calculating jet cross-sections
  in NLO QCD}, Nucl. Phys. B 485 (1997) 291--419, [Erratum: Nucl.\ Phys.\ B 510
  (1998) 503--504].
\newblock \href {http://arxiv.org/abs/hep-ph/9605323}
  {\path{arXiv:hep-ph/9605323}}, \href
  {https://doi.org/10.1016/S0550-3213(96)00589-5}
  {\path{doi:10.1016/S0550-3213(96)00589-5}}.

\bibitem{Dittmaier:1999mb}
S.~Dittmaier, {A general approach to photon radiation off fermions}, Nucl.
  Phys. B 565 (2000) 69--122.
\newblock \href {http://arxiv.org/abs/hep-ph/9904440}
  {\path{arXiv:hep-ph/9904440}}, \href
  {https://doi.org/10.1016/S0550-3213(99)00563-5}
  {\path{doi:10.1016/S0550-3213(99)00563-5}}.

\bibitem{Dittmaier:2008md}
S.~Dittmaier, A.~Kabelschacht, T.~Kasprzik, {Polarized QED splittings of
  massive fermions and dipole subtraction for non-collinear-safe observables},
  Nucl. Phys. B800 (2008) 146--189.
\newblock \href {http://arxiv.org/abs/0802.1405} {\path{arXiv:0802.1405}},
  \href {https://doi.org/10.1016/j.nuclphysb.2008.03.010}
  {\path{doi:10.1016/j.nuclphysb.2008.03.010}}.

\bibitem{Frixione:2007vw}
S.~Frixione, P.~Nason, C.~Oleari, {Matching NLO QCD computations with Parton
  Shower simulations: the POWHEG method}, JHEP 11 (2007) 070.
\newblock \href {http://arxiv.org/abs/0709.2092} {\path{arXiv:0709.2092}},
  \href {https://doi.org/10.1088/1126-6708/2007/11/070}
  {\path{doi:10.1088/1126-6708/2007/11/070}}.

\bibitem{Frederix:2009yq}
R.~Frederix, S.~Frixione, F.~Maltoni, T.~Stelzer, {Automation of
  next-to-leading order computations in QCD: The FKS subtraction}, JHEP 10
  (2009) 003.
\newblock \href {http://arxiv.org/abs/0908.4272} {\path{arXiv:0908.4272}},
  \href {https://doi.org/10.1088/1126-6708/2009/10/003}
  {\path{doi:10.1088/1126-6708/2009/10/003}}.

\bibitem{Melnikov:1993np}
K.~Melnikov, O.~I. Yakovlev, {Top near threshold: All $\alpha_S$ corrections
  are trivial}, Phys. Lett. B 324 (1994) 217--223.
\newblock \href {http://arxiv.org/abs/hep-ph/9302311}
  {\path{arXiv:hep-ph/9302311}}, \href
  {https://doi.org/10.1016/0370-2693(94)90410-3}
  {\path{doi:10.1016/0370-2693(94)90410-3}}.

\bibitem{Denner:2012yc}
A.~Denner, S.~Dittmaier, S.~Kallweit, S.~Pozzorini, {NLO QCD corrections to
  off-shell top-antitop production with leptonic decays at hadron colliders},
  JHEP 10 (2012) 110.
\newblock \href {http://arxiv.org/abs/1207.5018} {\path{arXiv:1207.5018}},
  \href {https://doi.org/10.1007/JHEP10(2012)110}
  {\path{doi:10.1007/JHEP10(2012)110}}.

\bibitem{Heinrich:2017bqp}
G.~Heinrich, et~al., {NLO and off-shell effects in top quark mass
  determinations}, JHEP 07 (2018) 129.
\newblock \href {http://arxiv.org/abs/1709.08615} {\path{arXiv:1709.08615}},
  \href {https://doi.org/10.1007/JHEP07(2018)129}
  {\path{doi:10.1007/JHEP07(2018)129}}.

\bibitem{Blondel:2019yqr}
N.~Alipour~Tehrani, et~al., {FCC-ee: Your Questions Answered}, in: A.~Blondel,
  P.~Janot (Eds.), {CERN Council Open Symposium on the Update of European
  Strategy for Particle Physics}, 2019.
\newblock \href {http://arxiv.org/abs/1906.02693} {\path{arXiv:1906.02693}}.

\bibitem{Groote:2010zf}
S.~Groote, J.~G. K{\"o}rner, B.~Melic, S.~Prelovsek, {A survey of top quark
  polarization at a polarized linear $e^+ e^-$ collider}, Phys. Rev. D 83
  (2011) 054018.
\newblock \href {http://arxiv.org/abs/1012.4600} {\path{arXiv:1012.4600}},
  \href {https://doi.org/10.1103/PhysRevD.83.054018}
  {\path{doi:10.1103/PhysRevD.83.054018}}.

\bibitem{Alwall:2014hca}
J.~Alwall, et~al., {The automated computation of tree-level and next-to-leading
  order differential cross sections, and their matching to parton shower
  simulations}, JHEP 07 (2014) 079.
\newblock \href {http://arxiv.org/abs/1405.0301} {\path{arXiv:1405.0301}},
  \href {https://doi.org/10.1007/JHEP07(2014)079}
  {\path{doi:10.1007/JHEP07(2014)079}}.

\end{thebibliography}

\end{document}